\documentclass[amsfonts,12pt,thmsa,sw20bams]{article}
\usepackage{amsmath}
\usepackage{graphicx}

\input tcilatex
\begin{document}

\author{Choulakian, V. \and Universit\'{e} de Moncton, Canada \and email:
vartan.choulakian@umoncton.ca}
\title{Globally Homogenous Mixture Components and Local Heterogeneity of
Rank Data }
\date{August 2016}
\maketitle

\begin{abstract}
The traditional methods of finding mixture components of rank data are
mostly based on distance and latent class models; these models may exhibit
the phenomenon of masking of groups of small sizes; probably due to the
spherical nature of rank data. Our approach diverges from the traditional
methods; it is directional and uses a logical principle, the law of
contradiction. We discuss the concept of a mixture for rank data essentially
in terms of the notion of global homogeneity of its group components. Local
heterogeneities may appear once the group components of the mixture have
been discovered. This is done via the exploratory analysis of rank data by
taxicab correspondence analysis with the nega coding: If the first factor is
an affine function of the Borda count, then we say that the rank data are
globally homogenous, and local heterogeneities may appear on the consequent
factors; otherwise, the rank data either are globally homogenous with
outliers, or a mixture of globally homogenous groups. Also we introduce a
new coefficient of global homogeneity, GHC. GHC is based on the first
taxicab dispersion measure: it takes values between 0 and 100\%, so it is
easily interpretable. GHC measures the extent of crossing of scores of
voters between two or three blocks seriation of the items where the Borda
count statistic provides consensus ordering of the items on the first axis.
Examples are provided.

\end{abstract}

\section{\bf Introduction}

Rankings of the elements of a set is a common daily decision making
activity, such as, voting for a political candidate, choosing a consumer
product, etc. So there is a huge literature concerning the analysis and
interpretation of preference rankings data. However, if we trace back in
time, we find that de Borda (1781) was the first author, who outlined a
simple well thought method based on a solid argument. Borda, as a member of
the French Academy of Sciences, criticised the plurality method of choosing
a new academy member and suggested, what is known as the Borda count(BC)
rule, to fully rank order (seriate) the $d$ candidates based on the
preferences of the $n$ judges. BC has generated a large litterature, and
this paper makes full use of it as much needed.

Let $A=\{a_{1},a_{2},\ldots ,a_{d}\}$ denote a set of $d$
alternatives/candidates/items, and $V$ a set of $n$
voters/individuals/judges. In this paper we consider the linear
orderings/rankings/preferences, in which all $d$ objects are rank-ordered
according to their levels of desirability by the $n$ voters. We denote a
linear order by a sequence ${\bf s}=(a_{k_{1}}\succ a_{k_{2}}\succ \ldots
\succ a_{k_{d}})$, where $a_{k_{1}}\succ a_{k_{2}}$ means that the
alternative $a_{k_{1}}$ is preferred to the alternative $a_{k_{2}}.$ Let $%
{\bf \Psi }$ be the set of all linear orders on $A;$ the cardinality of $%
{\bf \Psi }$ is $d!$. A {\it voting profile} is a function $w$ from $V$ to $%
{\bf \Psi }$, that is, $w(V)={\bf \Psi }.$

We denote by $S_{d}$ the set of permutations of the elements of the set $%
\left\{ 0,1,2,...,d-1\right\} .$ The Borda score is a function $b$ from $%
{\bf \Psi }$ to $S_{d},$ where for a linear ordering ${\bf s}\in $ ${\bf %
\Psi }$, Borda assigned to the element $a_{k_{j}}$the score of $(d-j)$,
because $a_{k_{j}}$ is preferred to ($d-j)$ other alternatives, or
equivalently it is the $j$th most preferred alternative. We denote $b({\bf %
\Psi })={\bf R}$, where ${\bf R=(}r_{ij})$ is a matrix having $n$ rows and $%
d $ columns, and $r_{ij}$ designates the Borda score of the $i$th judge's
preference of the $j$th alternative. The average Borda score of the elements
of $A$ is ${\bf \beta }={\bf 1}_{n}^{\prime }{\bf R}/n,$ where ${\bf 1}_{n}$
is a column vector of $1$'s having $n$ coordinates. Borda's count rule (BC)
seriates/orders the $d$ elements of the set $A$ according to their average
scores: ${\bf \beta }(j)>{\bf \beta }(i)$ means alternative $j$ is preferred
to alternative $i$.

We define the reverse Borda score to be a function $\overline{b}$ from ${\bf %
\Psi }$ to $S_{d},$ where for a linear order ${\bf s}\in $ ${\bf \Psi }$, we
assign to the element $a_{k_{j}}$the score of $(j-1).$ We denote $\overline{b%
}({\bf \Psi })=\overline{{\bf R}}$, where $\overline{{\bf R}}{\bf =(}%
\overline{r}_{ij})$ is a matrix having $n$ rows and $d$ columns, and $%
\overline{r}_{ij}$ designates the reverse Borda score of the $i$th judge's
preference of the $j$th alternative. The average reverse Borda score of the
elements of $A$ is $\overline{{\bf \beta }}={\bf 1}_{n}^{\prime }\overline{%
{\bf R}}/n.$

We note that
\[
{\bf R+}\overline{{\bf R}}=(d-1){\bf 1}_{n}{\bf 1}_{d}^{\prime }
\]%
and
\[
{\bf \beta +}\overline{{\bf \beta }}=(d-1){\bf 1}_{d}^{\prime }.
\]

\subsection{\it Example 1: Croon's political goals data}

This example has two aims: First to make the notation clear; second to show
that traditional well established methods for rank data, such as distance
and latent class based, may mask groups of small sizes in mixture models.
Table 1 introduces a well known data set first analyzed by Croon (1989); the
data derive from a german survey of 2262 rankings of four political items
concerning Inglehart (1977)'s theory of postmodernism. The four items are: ($%
A$) maintaining order in the nation; ($B$) giving people more to say in
important government decisions; ($C$) fighting rising prices; ($D$)
protecting freedom of speech. Inglehart advanced the thesis that there is a
shift in political culture in Europe; that is, some younger Europeans have
different political values than their fathers: He named the elder Europeans
as materialists, because after the first and second world wars, they valued
mostly material security item ($C$) and domestic order item ($A$); while he
named some of the younger generation as postmaterialists, because they
valued much more human rights and political liberties item ($D$) and
democracy item ($B$). So in this example $d=4$, $n=2262$, and the voting
profile is displayed in the first two columns of Table 1; similarly, Table 1
displays the Borda scores ${\bf R}$ and the reverse Borda scores $\overline{%
{\bf R}}$. The average BC score ${\bf \beta }$ and the average reverse BC
score $\overline{{\bf \beta }}$ show that, the 2262 voters generally rank
materialist items $\left\{ A,C\right\} \ \succ $\ postmaterialist items $%
\left\{ B,D\right\} $.

\begin{tabular}{|l|c|r|r|r|r|r|r|r|r|}
\multicolumn{10}{l}{\bf Table 1: Political Goals Data set of 2262 rankings
of four} \\
\multicolumn{10}{l}{\ \ \ \ \ \ \ {\bf \ \ \ items concerning Inglehart's
theory of postmodernism.}} \\ \hline
Item & observed & \multicolumn{4}{||l|}{Borda scores ${\bf R}$} &
\multicolumn{4}{||l|}{reverse Borda scores$\ \overline{{\bf R}}$} \\
\cline{3-10}
ordering & \multicolumn{1}{|l|}{frequency} & \multicolumn{1}{||r|}{A} & B & C
& D & \multicolumn{1}{||r|}{A} & B & C & D \\ \hline
A$\succ $B$\succ $C$\succ $D & \multicolumn{1}{|r|}{137} &
\multicolumn{1}{||r|}{3} & 2 & 1 & 0 & \multicolumn{1}{||r|}{0} & 1 & 2 & 3
\\
A$\succ $B$\succ $D$\succ $C & \multicolumn{1}{|r|}{29} &
\multicolumn{1}{||r|}{3} & 2 & 0 & 1 & \multicolumn{1}{||r|}{0} & 1 & 3 & 2
\\
A$\succ $C$\succ $B$\succ $D & \multicolumn{1}{|r|}{309} &
\multicolumn{1}{||r|}{3} & 1 & 2 & 0 & \multicolumn{1}{||r|}{0} & 2 & 1 & 3
\\
A$\succ $C$\succ $D$\succ $B & \multicolumn{1}{|r|}{255} &
\multicolumn{1}{||r|}{3} & 0 & 2 & 1 & \multicolumn{1}{||r|}{0} & 3 & 1 & 2
\\ \hline
A$\succ $D$\succ $B$\succ $C & \multicolumn{1}{|r|}{52} &
\multicolumn{1}{||r|}{3} & 1 & 0 & 2 & \multicolumn{1}{||r|}{0} & 2 & 3 & 1
\\
A$\succ $D$\succ $C$\succ $B & \multicolumn{1}{|r|}{93} &
\multicolumn{1}{||r|}{3} & 0 & 1 & 2 & \multicolumn{1}{||r|}{0} & 3 & 2 & 1
\\
B$\succ $A$\succ $C$\succ $D & \multicolumn{1}{|r|}{48} &
\multicolumn{1}{||r|}{2} & 3 & 1 & 0 & \multicolumn{1}{||r|}{1} & 0 & 2 & 3
\\
B$\succ $A$\succ $D$\succ $C & \multicolumn{1}{|r|}{23} &
\multicolumn{1}{||r|}{2} & 3 & 0 & 1 & \multicolumn{1}{||r|}{1} & 0 & 3 & 2
\\ \hline
B$\succ $C$\succ $A$\succ $D & \multicolumn{1}{|r|}{61} &
\multicolumn{1}{||r|}{1} & 3 & 2 & 0 & \multicolumn{1}{||r|}{2} & 0 & 1 & 3
\\
B$\succ $C$\succ $D$\succ $A & \multicolumn{1}{|r|}{55} &
\multicolumn{1}{||r|}{0} & 3 & 2 & 1 & \multicolumn{1}{||r|}{3} & 0 & 1 & 2
\\
B$\succ $D$\succ $A$\succ $C & \multicolumn{1}{|r|}{33} &
\multicolumn{1}{||r|}{1} & 3 & 0 & 2 & \multicolumn{1}{||r|}{2} & 0 & 3 & 1
\\
B$\succ $D$\succ $C$\succ $A & \multicolumn{1}{|r|}{59} &
\multicolumn{1}{||r|}{0} & 3 & 1 & 2 & \multicolumn{1}{||r|}{3} & 0 & 2 & 1
\\ \hline
C$\succ $A$\succ $B$\succ $D & \multicolumn{1}{|r|}{330} &
\multicolumn{1}{||r|}{2} & 1 & 3 & 0 & \multicolumn{1}{||r|}{1} & 2 & 0 & 3
\\
C$\succ $A$\succ $D$\succ $B & \multicolumn{1}{|r|}{294} &
\multicolumn{1}{||r|}{2} & 0 & 3 & 1 & \multicolumn{1}{||r|}{1} & 3 & 0 & 2
\\
C$\succ $B$\succ $A$\succ $D & \multicolumn{1}{|r|}{117} &
\multicolumn{1}{||r|}{1} & 2 & 3 & 0 & \multicolumn{1}{||r|}{2} & 1 & 0 & 3
\\
C$\succ $B$\succ $D$\succ $A & \multicolumn{1}{|r|}{69} &
\multicolumn{1}{||r|}{0} & 2 & 3 & 1 & \multicolumn{1}{||r|}{3} & 1 & 0 & 2
\\ \hline
C$\succ $D$\succ $A$\succ $B & \multicolumn{1}{|r|}{70} &
\multicolumn{1}{||r|}{1} & 0 & 3 & 2 & \multicolumn{1}{||r|}{2} & 3 & 0 & 1
\\
C$\succ $D$\succ $B$\succ $A & \multicolumn{1}{|r|}{34} &
\multicolumn{1}{||r|}{0} & 1 & 3 & 2 & \multicolumn{1}{||r|}{3} & 2 & 0 & 1
\\
D$\succ $A$\succ $B$\succ $C & \multicolumn{1}{|r|}{21} &
\multicolumn{1}{||r|}{2} & 1 & 0 & 3 & \multicolumn{1}{||r|}{1} & 2 & 3 & 0
\\
D$\succ $A$\succ $C$\succ $B & \multicolumn{1}{|r|}{30} &
\multicolumn{1}{||r|}{2} & 0 & 1 & 3 & \multicolumn{1}{||r|}{1} & 3 & 2 & 0
\\ \hline
D$\succ $B$\succ $A$\succ $C & \multicolumn{1}{|r|}{29} &
\multicolumn{1}{||r|}{1} & 2 & 0 & 3 & \multicolumn{1}{||r|}{2} & 1 & 3 & 0
\\ \hline
D$\succ $B$\succ $C$\succ $A & \multicolumn{1}{|r|}{52} &
\multicolumn{1}{||r|}{0} & 2 & 1 & 3 & \multicolumn{1}{||r|}{3} & 1 & 2 & 0
\\ \hline
D$\succ $C$\succ $A$\succ $B & \multicolumn{1}{|r|}{35} &
\multicolumn{1}{||r|}{1} & 0 & 2 & 3 & \multicolumn{1}{||r|}{2} & 3 & 1 & 0
\\ \hline
D$\succ $C$\succ $B$\succ $A & \multicolumn{1}{|r|}{27} &
\multicolumn{1}{||r|}{0} & 1 & 2 & 3 & \multicolumn{1}{||r|}{3} & 2 & 1 & 0
\\ \hline
${\bf \beta }$ &  & \multicolumn{1}{|l|}{1.97} & 1.10 & 2.05 & 0.88 &
\multicolumn{1}{|l|}{} &  &  &  \\ \hline
$\overline{{\bf \beta }}$ &  & \multicolumn{1}{|l|}{} &  &  &  &
\multicolumn{1}{|l|}{1.03} & 1.90 & 0.95 & 2.12 \\ \hline
\end{tabular}

Table 2 provides a statistical summary of four methods of data analysis of
Table 1. The first method suggested by Inglehart is {\it deductive and
supervised}; it opposes to the other three methods, which are {\it %
inductive, unsupervised and aim to validate Inglehart's theory of
postmaterialism,} see also Moors and Vermunt (2007). The other three methods
are mixture models and they attempt to see if this data set confirms
Inglehart's theory of postmaterialism. The first one is by Croon (1989), who
used a stochastic utility (SU) based latent class model; the second one by
Lee and Yu (2012), who used a weighted distance-based Footrule mixture
model; and the third one is based on taxicab correspondence analysis (TCA),
which is the topic of this paper. Here, we provide some details on the
statistics displayed in Table 2.

a) Inglehart (1977) {\it apriori} classified the respondents into three
groups: materialists, postmaterialists and mixed. His method of
classification is based on partial rankings based on the first two preferred
choices. Here, we discuss each group separately.

Materialists are defined by their response patterns $(A\succ C\succ B\succ
D, $ $A\succ C\succ D\succ B,$ $C\succ A\succ B\succ D,$ $C\succ A\succ
D\succ B)$, where the pair of materialist items $\left\{ A,C\right\} $ are
always ranked above the pair of postmaterialist items $\left\{ B,D\right\} $%
; they make $52.52\%$ of the voters. In the ideal case we expect to have the
average BC scores for the four items to be: ${\bf \beta }_{ideal}(A)={\bf %
\beta }_{ideal}(C)=2.5$ and ${\bf \beta }_{ideal}(B)={\bf \beta }%
_{ideal}(D)=0.5;$ the corresponding observed values, displayed in Table 2,
are (very near to the ideal ones): ${\bf \beta }(A)=2.4747\simeq {\bf \beta }%
(C)=2.5253$ and ${\bf \beta }(B)=0.5379\simeq {\bf \beta }(D)=0.4621.$

Postmaterialists are defined by response paterns ($B\succ D\succ A\succ C,$ $%
B\succ D\succ C\succ A,$ $D\succ B\succ A\succ C,$ $D\succ B\succ C\succ A$%
), where the pair of postmaterialist items $\left\{ B,D\right\} $ are always
ranked above the pair of materialist items $\left\{ A,C\right\} $; they make
$7.65\%$ of the voters. The comparison of ideal and observed average BC
scores, displayed in Table 2, show that: ${\bf \beta }_{ideal}(B)={\bf \beta
}_{ideal}(D)=2.5$ is very near to ${\bf \beta }(B)=2.5318\simeq {\bf \beta }%
(D)=2.4682,$ while ${\bf \beta }_{ideal}(A)={\bf \beta }_{ideal}(C)=0.5$ is
somewhat near to ${\bf \beta }(A)=0.3584,$ ${\bf \beta }(C)=0.6416$.

The last group is named 'mixed' by Inglehart and is composed of the
remaining sixteen response patterns; they make $39.83\%$ of the voters. In
the ideal case we expect to have the average BC scores for the four items to
be: ${\bf \beta }_{ideal}(A)={\bf \beta }_{ideal}(C)=$ ${\bf \beta }%
_{ideal}(B)={\bf \beta }_{ideal}(D)=1.5;$ the corresponding observed values,
displayed in Table 2, are (somewhat near to the ideal ones): ${\bf \beta }%
(A)=1.6204\simeq $ ${\bf \beta }(C)=1.7026\simeq $ ${\bf \beta }(B)=1.5527$
and ${\bf \beta }(D)=1.1243.$

Furthermore, based on the global homogeneity coefficient GHC in \%: $%
GHC(materialists)=100\%$ and $GHC(postmaterialist)=100\%.$ Inglehart's mixed
group is not globally homogenous; that is why we did not calculate its GHC
index. The development of the GHC index and its interpretation will be done
in section 3.

It is important to note that, the underlying hypothetical
conceptual-structural model for this data is a mixture composed of three
specific groups (materialist, postmaterialist and mixed), which are
explicitly characterized by Inglehart.

b,c) Given that, Croon's SU model and Lee and Yu's weighted distance-based
Footrule mixture model produced globally very similar groups, we present
them together. A summary of Croon's analysis can also be found in Skrondal
and Rabe-Hesketh (2004, p.404-406), Lee and Yu (2012) and in Alvo and Yu
(2014, p.228-232). Sections b and c of Table 2 are taken from Alvo and Yu
(2014, p. 230), who present a summary and a comparison of results from Croon
(1989) and Lee and Yu (2012). The interpretation of the estimated parameters
of the SU model in Table 2 is similar to the average Borda score: for each
group the score $s(item)$ shows the intensity of the preference for that
item in an increasing order. There are two kinds of estimated parameters in
Lee and Yu's weighted distance-based Footrule mixture model: the modal
response pattern for each group is shown in the last column; and the weight
of an item $w(item)$, which reflects our confidence in the ranked position
of the item in the modal response pattern, the higher value representing
higher confidence. Both methods find a mixture of three groups similar in
contents: the first two groups represent materialists with $35.2\%$ and $%
44.1\%$ of the voters for the weighted footrule mixture model, and $44.9\%$
and $32.6\%$ of the voters for the SU mixture model; and the third group
represents postmaterialists with $20.8\%$ of the voters for the weighted
footrule mixture model, and $22.5\%$ for the SU mixture model. Lee and Yu
(2012)'s conclusion is:\textquotedblright\ {\it Based on our grouping, we
may conclude that Inglehart's theory is not appropriate in Germany}%
\textquotedblright . This assertion shows that, the two well established
traditional methods masked the existence of the mixed group as put forth by
Inglehart.

d) Our approach, based on taxicab correspondence analysis (TCA), which is an
L$_{1}$ variant of correspondence analysis (CA), discovers a mixture of
three globally homogenous groups as advocated by Inglehart: Materialists
with $70.95\%$ of the voters, postmaterialists with $20.07\%$, and mixed
with $7.65\%$ of the voters. Furthermore, there is an outlier response
pattern ($D\succ A\succ C\succ B$) representing $1.33\%$ of the voters. So
contrary to Lee and Yu (2012)'s assertion, {\it our results validates
Inglehart's theory of postmodernism for this data set}. Probably, this is
due mainly to the fact that TCA is a directional method specially useful for
spherical data: Rank data with all its permutations is spherical by nature (
graphically, it is represented by a permutahedron; see Marden (2005, Figure
2.4, page 11) or Benz\'{e}cri (1980, p.303). Furthermore, based on the
global homogeneity coefficient GHC in \%: $GHC(materialists)=87.01\%$, $%
GHC(postmaterialist)=57.60\%,$ and $GHC(mixed)=72.82\%.$ We see that the
materialist voters form much more globally homogenous group than the voters
in the mixed group; and the voters in mixed group are much more homogenous
than the voters in the postmaterialist group. Furthermore, our analysis
clearly shows why the postmaterialists (they have three 'poles of
attractions' as defined by Marden (1995, ch. 2) or Benz\'{e}cri (1966,
1980)) are much more heterogenous than the materialists (they have two poles
of attractions). More details on the local heterogeneities of each group
will be presented later on in section 4.\bigskip

\begin{tabular}{|l|r||r|r|r|r|r}
\multicolumn{7}{l}{\bf Table 2: A summary of results derived from four
methods of analysis} \\
\multicolumn{7}{l}{\ \ \ \ \ \ \ \ \ \ \ \ \ \ \ {\bf of Political Goals\
Data set.}} \\ \hline
\multicolumn{7}{|l||}{\it a) Inglehart's a priori classification} \\ \hline
Group & sample\% & ${\bf \beta }(A)$ & ${\bf \beta }(B)$ & ${\bf \beta }(C)$
& ${\bf \beta }(D)$ & \multicolumn{1}{||r||}{$GHC(\%)$} \\ \hline
materialist & $52.52$ & $2.4747$ & $0.5379$ & $2.5253$ & $0.4621$ &
\multicolumn{1}{||r||}{$100$} \\
postmaterialist & $7.65$ & $0.3584$ & $2.5318$ & $0.6416$ & $2.4682$ &
\multicolumn{1}{||r||}{$100$} \\
mixed & $39.83$ & $1.6204$ & $1.5527$ & $1.7026$ & $1.1243$ &
\multicolumn{1}{||r||}{} \\ \hline
\multicolumn{7}{|l||}{\it b) Croon's SU mixture model} \\ \hline
Group & sample\% & $s(A)$ & $s(B)$ & $s(C)$ & $s(D)$ & \multicolumn{1}{||r}{}
\\ \cline{1-6}
materialist 1 & $44.9$ & $0.590$ & $-1.071$ & $1.730$ & $-1.249$ &
\multicolumn{1}{||r}{} \\
materialist 2 & $32.6$ & $1.990$ & $-0.920$ & $0.060$ & $-1.130$ &
\multicolumn{1}{||r}{} \\
postmaterialist & $22.5$ & $-0.691$ & $0.630$ & $-0.010$ & $0.071$ &
\multicolumn{1}{||r}{} \\ \hline
\multicolumn{7}{|l||}{\it c) Lee and Yu's weighted distance-based Footrule
mixture model} \\ \hline
Group & sample\% & $w1$ & $w2$ & $w3$ & \multicolumn{1}{|r||}{$w4$} &
\multicolumn{1}{|l||}{Modal ordering of items} \\ \hline
materialist 1 & $35.2$ & $2.030$ & $1.234$ & $0$ & \multicolumn{1}{|r||}{$%
0.191$} & \multicolumn{1}{|r||}{$C\succ A\succ B\succ D$} \\
materialist 2 & $44.1$ & $1.348$ & $0.917$ & $0.107$ & $0.104$ &
\multicolumn{1}{||r||}{$A\succ C\succ B\succ D$} \\
post-materialist & $20.8$ & $0.314$ & $0$ & $0.151$ & \multicolumn{1}{|r||}{$%
0.552$} & \multicolumn{1}{|r||}{$B\succ D\succ C\succ A$} \\ \hline
\multicolumn{7}{|l||}{\it d) Mixtures by taxicab correspondence analysis
with nega coding} \\ \hline
Group & sample\% & ${\bf \beta }(A)$ & ${\bf \beta }(B)$ & ${\bf \beta }(C)$
& ${\bf \beta }(D)$ & \multicolumn{1}{||r||}{$GHC(\%)$} \\ \hline
materialist & $70.95$ & $2.38$ & $0.72$ & $2.36$ & $0.54$ &
\multicolumn{1}{||r||}{$87.01$} \\
postmaterialist & $20.07$ & $0.348$ & $2.17$ & $1.71$ & $1.77$ &
\multicolumn{1}{||r||}{$57.60$} \\
mixed & $7.65$ & $2.46$ & $1.99$ & $0.28$ & $1.23$ & \multicolumn{1}{||r||}{$%
72.82$} \\
Outlier & $1.33$ & $2$ & $0$ & $1$ & $3$ & \multicolumn{1}{||r||}{} \\ \hline
\end{tabular}

\subsection{Organisation of this paper}

The traditional methods of finding mixture components of rank data are
mostly based on distance and latent class models; these models may mask
groups of small sizes; probably due to the spherical nature of rank data. In
this paper, our apparoach diverges from the traditional methods, because we
discuss the concept of a mixture for rank data essentially in terms of its
globally homogenous group components. We use the law of contradiction to
identify globally homogenous components. For instance, by TCA we were able
to discover that the data set in Table 1 is a mixture of three globally
homogenous group components (materialist, postmaterialist and mixed);
furthermore, each group component can be summarized by its average Borda
Count (BC) score as its consensus ranking; this is the first step in our
procedure. In the second step, we look at local heterogeneities if there are
any, given the globally homogenous component. This two step procedure
produces finer visualization of rank data; it is done via the exploratory
analysis of rank data by taxicab correspondence analysis with the nega
coding. Also we introduce a new coefficient of global homogeneity, GHC. GHC
is based on the first taxicab dispersion measure: it takes values between 0
and 100\%, so it is easily interpretable. GHC measures the extent of
crossing of scores of voters between 2 or 3 blocks seriation of the items
where the Borda count statistic provides consensus ordering of the items on
the first axis. Furthermore, to our knowledge, this is the first time that a
tangible method has been proposed that identifies explicitly outliers in a
rank data: neither the recently written monograph by Alvo and Yu (2014), nor
the much cited monograph of Marden (1995) discuss the important problem of
identification of outliers in rank data. We mention two publicly available
written packages in R, that we used, {\it RankClustr} by Jacques, Grimonprez
and Biernacki (2014), and {\it Pmr }(probability models for ranking data) by
Lee and Yu (2013).

The contents of this paper are organized as follows: Section 2 reviews the
TCA approach for rank data; section 3 develops the new global homogeneity
coefficient GHC; section 4 presents the analysis of some well known rank
data sets by TCA; and finally in section 5 we conclude with some remarks.

We just want to mention that there is a large litterature in social choice
theory or social welfare theory studying the properties of the BC. Here, we
mention some important contributions according to our personal readings.
Young (1974) presents a set of four axioms that characterize uniquely BC;
see also among others, Saary (1990a) and Marchant (1998). Saari (1990b)
distinguishes two levels of susceptiblity of manipulation of voting
theories: '{\it macro}'- where a large percentage of voters-, and '{\it micro%
}'- where a small percentage of voters - attempt to change the results of
the elections. In data analysis, a macro manipulation is equivalent to the
existence of a mixture of groups of voters. While, a micro manipulation is
equivalent to the existence of few outliers in the globally homogenous set
of voters $V$. Further, Saari concludes that among all positional voting
systems, BC is the least susceptible to micro manipulation; this assertion
seems fully true in this paper. Saari (1999) proves that BC is the only
positional voting method that satisfies the property of {\it Reversal
Symmetry, }which states that if everyone reverses all their preferences,
then the final outcome should also be reversed. This property plays an
important role in the nega coding of a rank data set before the application
of TCA. Choulakian (2014) incorporates the BC to interpret the first
principal factor of taxicab correspondence analysis (TCA) of a nega coded
rank data, see Theorem 1 in the next section. Additionally, this essay
further extends and complements the ideas of global homogeneity and local
heterogeneities for rank data.

\section{Taxicab correspondence analysis of nega coded rank data}

Results of this section are taken from Choulakian (2006, 2014). We start
with an overview of TCA of a contingency table; then review the
corresponding results concerning rank data.

\subsection{\bf Taxicab Correspondence analysis: An overview}

Let ${\bf X=(}x_{ij})$ be a contingency table cross-classifying two nominal
variables with $I$ rows and $J$ columns, and ${\bf P=X/}x_{\ast \ast }$ be
the associated correspondence matrix with elements $p_{ij},$ where $x_{\ast
\ast }=\sum_{j=1}^{J}\sum_{i=1}^{I}x_{ij}$ is the sample size. We define as
usual $p_{i\ast }=\sum_{j=1}^{J}p_{ij}$ , $p_{\ast j}=\sum_{i=1}^{I}p_{ij},$
the vector ${\bf r=(}p_{i\ast })\in{\mathbf{R}}^{I},$ the vector ${\bf c=(}p_{\ast j})\in\mathbf{R}^{J}$, and ${\bf D}_{r}=Diag({\bf r})$ a diagonal matrix having diagonal
elements $p_{i\ast },$ and similarly ${\bf D}_{c}=Diag({\bf c}).$ Let $%
k=rank({\bf P)-}1.$

In TCA we compute the following quadruplets for the two spaces, for $\alpha
=1,...,k$: $({\bf u}_{\alpha },{\bf b}_{\alpha },{\bf g}_{\alpha },\lambda
_{\alpha })$ in the row space of ${\bf P}$ and $({\bf v}_{\alpha },{\bf a}%
_{\alpha },{\bf f}_{\alpha },\lambda _{\alpha })$ in the column space of $%
{\bf P}$. Given that in CA and TCA, the row and column spaces are dual to
each other, we name the pair of vectors (${\bf u}_{\alpha }$ and ${\bf v}%
_{\alpha })$ $\alpha $th principal axes, the pair (${\bf a}_{\alpha }$ and $%
{\bf b}_{\alpha })$ $\alpha $th basic vectors of coordinates, the pair ($%
{\bf f}_{\alpha }$ and ${\bf g}_{\alpha })$ $\alpha $th vectors containg TCA
factor scores, and the nonnegative scalar $\lambda _{\alpha }$ the $\alpha $%
th TCA dispersion measure. The relations among the seven terms will be
described in the next two subsections.

TCA is computed in 2 steps: In the first step we compute the taxicab
singular value decomposition (TSVD) of ${\bf P}$ as a function of $({\bf a}%
_{\alpha },{\bf b}_{\alpha },\lambda _{\alpha })$ for $\alpha =1,...,k{\bf ,}
$ which is a stepwise matrix decomposition method based on a particular
matrix norm, see below equation (3). In the 2nd step, we reweight the pair
of basic vectors $({\bf a}_{\alpha },{\bf b}_{\alpha })$ by respective
weights of the columns, ${\bf D}_{c},$ and the rows, ${\bf D}_{r},$ to
obtain the vectors of factor scores $({\bf f}_{\alpha },{\bf g}_{\alpha })$
for $\alpha =1,...,k$.

\subsection{Taxicab singular value decomposition}

Let ${\bf P}^{(\alpha )}$ be the residual data matrix at the $\alpha $th
iteration, where, ${\bf P}^{(0)}={\bf P}$ for\ $\alpha =0.$ TSVD consists of
maximizing the $L_{1}$ norm of the linear combination of the columns of the
matrix ${\bf P}^{(\alpha )}$ subject to $L_{\infty }$ norm constraint, where
the $L_{1}$ norm of \ a vector ${\bf v}=(v_{1},...,v_{m})^{\prime }$ is
defined to be $\left\vert \left\vert {\bf v}\right\vert \right\vert
_{1}=\sum_{i=1}^{m}\left\vert v_{i}\right\vert $ $\ $and $\ \left\vert
\left\vert {\bf v}\right\vert \right\vert _{\infty }=\max_{i}\left\vert
v_{i}\right\vert $ is the $L_{\infty }$ norm; more precisely, it is based on
the following optimization problem%
\begin{equation}
max\left\vert \left\vert {\bf P}^{(\alpha )}{\bf u}\right\vert \right\vert
_{1}\ \ \text{subject to }\left\vert \left\vert {\bf u}\right\vert
\right\vert _{\infty }=1{\bf ;}
\end{equation}%
or equivalently, it can also be described as maximization of the $L_{1}$
norm of the linear combination of the rows of the matrix ${\bf P}^{(\alpha
)} $%
\begin{equation}
max\left\vert \left\vert {\bf P}^{(\alpha )\prime }{\bf v}\right\vert
\right\vert _{1}\ \ \text{subject to }\left\vert \left\vert {\bf v}%
\right\vert \right\vert _{\infty }=1{\bf .}
\end{equation}%
Equation (1) is the dual of (2), and they can be reexpressed as matrix
operator norms
\begin{eqnarray}
\lambda _{\alpha } &=&\max_{{\bf u\in\mathbf{R}}^{J}}\frac{\left\vert \left\vert {\bf P}^{(\alpha )}{\bf u}\right\vert
\right\vert _{1}}{\left\vert \left\vert {\bf u}\right\vert \right\vert
_{\infty }},  \nonumber \\
&=&\max {\bf ||P^{(\alpha )}u||}_{1}\ \ \text{subject to }{\bf u}\in \left\{
-1,+1\right\} ^{J},  \nonumber \\
&=&\max_{{\bf v\in\mathbf{R}}^{I}}\frac{\left\vert \left\vert {\bf P}^{(\alpha )\prime }{\bf v}%
\right\vert \right\vert _{1}}{\left\vert \left\vert {\bf v}\right\vert
\right\vert _{\infty }}, \\
&=&\max_{{\bf v}}\left\vert \left\vert {\bf P}^{(\alpha )\prime }{\bf v}%
\right\vert \right\vert _{1}\text{ such to }{\bf v}\in \left\{ -1,+1\right\}
^{I}  \nonumber \\
&=&\max_{{\bf u\in\mathbf{R}}^{J},{\bf v\in\mathbf{R}}^{I}}\frac{{\bf v}^{\prime }{\bf P}^{(\alpha )}{\bf u}}{\left\vert
\left\vert {\bf u}\right\vert \right\vert _{\infty }\left\vert \left\vert
{\bf v}\right\vert \right\vert _{\infty }},  \nonumber
\end{eqnarray}%
which is a well known and much discussed matrix norm related to the
Grothendieck problem; the inequality in Theorem 2 section 3 of this paper
sheds further insight into Grothendieck's theorem; see Pisier (2012) for a
comprehensive and interesting history of Grothendieck's theorem with its
many variants.

Equation (3) characterizes the robustness of the method, in the sense that,
the weights affected to the columns (similarly to the rows by duality) are
uniform $\pm 1.$ The $\alpha $th principal axes, ${\bf u}_{\alpha }$ and $%
{\bf v}_{\alpha },$ are computed by%
\begin{equation}
{\bf u}_{\alpha }=\arg \max_{{\bf u}}\left\vert \left\vert {\bf P}^{(\alpha
)}{\bf u}\right\vert \right\vert _{1}\text{ such that }\left\vert \left\vert
{\bf u}\right\vert \right\vert _{\infty }=1,
\end{equation}%
and%
\begin{equation}
{\bf v}_{\alpha }=\arg \max_{{\bf v}}\left\vert \left\vert {\bf P}^{(\alpha
)\prime }{\bf v}\right\vert \right\vert _{1}\text{ such that }\left\vert
\left\vert {\bf v}\right\vert \right\vert _{\infty }=1.
\end{equation}%
It is evident that for $\alpha =0$%
\begin{equation}
{\bf u}_{0}={\bf 1}_{J}\text{ and }{\bf v}_{0}={\bf 1}_{I},
\end{equation}%
where ${\bf 1}_{J}$ represents a column vector of ones of length $J.$ The
two principal axes ${\bf u}_{0}$ and ${\bf v}_{0}$ are named trivial, and
they are used only to center the rows and the columns of ${\bf P}$.

Let ${\bf a}_{\alpha }$ represent the $\alpha $th TSVD coordinates of the
rows of ${\bf P}^{(\alpha )}$ by projecting the rows of ${\bf P}^{(\alpha )}$
on the principal axis ${\bf u}_{\alpha }$, and ${\bf b}_{\alpha }$ represent
the $\alpha $th TSVD coordinates of the columns of ${\bf P}^{(\alpha )}$ by
projecting the columns of ${\bf P}^{(\alpha )}$ on the principal axis ${\bf v%
}_{\alpha }$. These are given by

\begin{equation}
{\bf a}_{\alpha }={\bf P}^{(\alpha )}{\bf u}_{\alpha }\text{ and }{\bf b}%
_{\alpha }={\bf P}^{(\alpha )\prime }{\bf v}_{\alpha };
\end{equation}%
and%
\begin{equation}
\left\vert \left\vert {\bf a}_{\alpha }\right\vert \right\vert _{1}={\bf v}%
_{\alpha }^{\prime }{\bf a}_{\alpha }=\left\vert \left\vert {\bf b}_{\alpha
}\right\vert \right\vert _{1}={\bf u}_{\alpha }^{\prime }{\bf b}_{\alpha
}=\lambda _{\alpha }.
\end{equation}%
In particular, by (6,7,8), we have for $\alpha =0$%
\[
{\bf a}_{0}={\bf r}\text{ ,\ }{\bf b}_{0}={\bf c}\text{\ \ \ and\ \ }\lambda
_{0}=1.
\]%
Equations (7) are named transition formulas, because ${\bf v}_{\alpha }$ and
${\bf a}_{\alpha },$ and , ${\bf u}_{\alpha }$ and ${\bf b}_{\alpha },$ are
related by
\begin{equation}
{\bf u}_{\alpha }=sgn({\bf b}_{\alpha })\text{ \ \ and \ \ }{\bf v}_{\alpha
}=sgn({\bf a}_{\alpha }),
\end{equation}%
where $sgn({\bf b}_{\alpha })=(sgn(b_{\alpha }(1)),...,sgn(b_{\alpha
}(J))^{\prime },$ and $sgn(b_{\alpha }(j))=1$ if $b_{\alpha }(j)>0,$ $%
sgn(b_{\alpha }(j))=-1$ otherwise.

To obtain the $(\alpha +1)$th TSVD row and column coordinates ${\bf a}%
_{\alpha +1}$ and ${\bf b}_{\alpha +1},$ and corresponding principal axes $%
{\bf u}_{\alpha +1}$ and ${\bf v}_{\alpha +1}$, we repeat the above
procedure on the residual dataset
\begin{equation}
{\bf P}^{(\alpha +1)}{\bf =P^{(\alpha )}-a}_{\alpha }{\bf b}_{\alpha
}^{\prime }/\lambda _{\alpha }.
\end{equation}%
We note that the $rank({\bf P}^{(\alpha +1)})\ {\bf =\ }rank({\bf P}%
^{(\alpha )}{\bf )-}1,$ because by (6) through (9)
\begin{equation}
{\bf P}^{(\alpha +1)}{\bf u}_{\alpha }={\bf 0}\text{ \ and }{\bf P}^{(\alpha
+1)^{\prime }}{\bf v}_{\alpha }={\bf 0;}
\end{equation}%
which, by induction, implies that for $\alpha =1,...,k$
\[
{\bf u}_{\beta }^{\prime }{\bf b}_{\alpha }=0\text{ \ and \ }{\bf v}_{\beta
}^{\prime }{\bf a}_{\alpha }=0\text{ \ for }\beta =0,...,\alpha -1;
\]%
and in particular we see that for $\alpha =1,...,k$
\begin{equation}
{\bf 1}_{J}^{\prime }{\bf b}_{\alpha }=0\text{ \ and \ }{\bf 1}_{I}^{\prime }%
{\bf a}_{\alpha }=0\text{ \ by (6);}
\end{equation}%
that is, the basic vectors ${\bf b}_{\alpha }$ and ${\bf a}_{\alpha }$ are
centered.

The data reconstitution formula for the correspondence matrix ${\bf P}$ as a
function of the basic vectors $({\bf a}_{\alpha },{\bf b}_{\alpha })$ for $%
\alpha =1,...,k\ $associated with the dispersion measures $\lambda _{\alpha
} $ is%
\begin{equation}
p_{ij}=p_{i.}p_{.j}+\sum_{\alpha =1}^{k}a_{\alpha }(i)b_{\alpha }(j)/\lambda
_{\alpha }.
\end{equation}%
In TCA of ${\bf P}$ both basic vectors ${\bf a}_{\alpha }$ and ${\bf b}%
_{\alpha }$ for$\ \ \alpha =1,...,k$ satisfy the equivariability property,
as a consequence of equations (8,12) , see Choulakian (2008a). This means
that ${\bf a}_{\alpha }$ and ${\bf b}_{\alpha }$ are balanced in the sense
that%
\begin{eqnarray}
\frac{\lambda _{\alpha }}{2} &=&\sum_{i}\left[ a_{\alpha }(i)|a_{\alpha
}(i)>0\right]  \nonumber \\
&=&-\sum_{i}\left[ a_{\alpha }(i)|a_{\alpha }(i)<0\right] \\
&=&\sum_{j}\left[ b_{\alpha }(j)|b_{\alpha }(j)>0\right]  \nonumber \\
&=&-\sum_{j}\left[ b_{\alpha }(j)|b_{\alpha }(j)<0\right] .  \nonumber
\end{eqnarray}

In TSVD, the optimization problems (3), (4) or (5) can be accomplished by
two algorithms. The first one is based on complete enumeration (3); this can
be applied, with the present state of desktop computing power, say, when $%
min(I,J)\simeq 25.$ The second one is based on iterating the transitional
formulas (7), (8) and (9), similar to Wold's (1966) NIPALS (nonlinear
iterative partial alternating least squares) algorithm, also named
criss-cross regression by Gabriel and Zamir (1979). The criss-cross
nonlinear algorithm can be summarized in the following way, where ${\bf b}$
is a starting value:

Step 1: ${\bf u=}sgn{\bf (b)}$, ${\bf a=P^{(\alpha )}u}$\ and $\lambda ({\bf %
a)}=\left\vert \left\vert {\bf a}\right\vert \right\vert _{1};$

Step 2: ${\bf v=}sgn{\bf (a),}$ ${\bf b=P}^{(\alpha )\prime }{\bf v}$\ and $%
\lambda ({\bf b)}=\left\vert \left\vert {\bf b}\right\vert \right\vert _{1};$

Step 3: If $\lambda ({\bf b)-}\lambda ({\bf a)>}0{\bf ,}$ go to Step 1;
otherwise, stop.

This is an ascent algorithm, see Choulakian (2016); that is, it increases
the value of the objective function $\lambda $ at each iteration. The
convergence of the algorithm is superlinear (very fast, at most two or three
iterations); however it could converge to a local maximum; so we restart the
algorithm $I$ times using each row of ${\bf P}^{(\alpha )}$ as a starting
value. The iterative algorithm is statistically consistent in the sense that
as the sample size increases there will be some observations in the
direction of the principal axes, so the algorithm will find the optimal
solution.

\subsection{Taxicab correspondence analysis}

A simple reweighting of the basic coordinates $({\bf a}_{\alpha },{\bf b}%
_{\alpha })$ for$\ \ \alpha =1,...,k$ produces the vectors $({\bf f}_{\alpha
},{\bf g}_{\alpha })$ that contain TCA factor scores of the rows and the
columns

\begin{equation}
{\bf f}_{\alpha }={\bf D}_{r}^{-1}{\bf a}_{\alpha }\text{ and }{\bf g}%
_{\alpha }={\bf D}_{c}^{-1}{\bf b}_{\alpha }\text{;}
\end{equation}%
and (8) becomes%
\begin{equation}
{\bf v}_{\alpha }^{\prime }{\bf D}_{r}{\bf f}_{\alpha }={\bf u}_{\alpha
}^{\prime }{\bf D}_{c}{\bf g}_{\alpha }=\lambda _{\alpha }.
\end{equation}

Similar to CA, TCA satisfies an important invariance property: columns (or
rows) with identical profiles (conditional probabilities) receive identical
factor scores. Moreover, merging of identical profiles does not change the
result of the data analysis: This is named the principle of equivalent
partitioning by Nishisato (1984); it includes the famous distributional
equivalence property of Benz\'{e}cri, which is satisfied by CA.

By (13 and 15), one gets the data reconstitution formula in TCA (the same
formula as in CA) for the correspondence matrix ${\bf P}$ as a function of
the factor coordinates $({\bf f}_{\alpha },{\bf g}_{\alpha })$ for $\alpha
=1,...,k\ $associated with the eigenvalues $\lambda _{\alpha }$
\begin{equation}
p_{ij}=p_{i.}p_{.j}\left[ 1+\sum_{\alpha =1}^{k}f_{\alpha }(i)g_{\alpha
}(j)/\lambda _{\alpha }\right] .
\end{equation}%
The visual maps are obtained by plotting the points $(f_{\alpha
}(i),f_{\beta }(i))$ for $i=1,...,I$ or $(g_{\alpha }(j),g_{\beta }(j))$ for
$j=1,...,J,$ for $\alpha \neq \beta .$

Correspondence analysis (CA) admits a chi-square distance interpretation
between profiles; there is no chi-square like distance in TCA. Fichet (2009)
described it as a general scoring method.

In the sequel we suppose that the theory of correspondence analysis (CA) is
known. The theory of CA can be found, among others, in Benz\'{e}cri (1973,
1992), Greenacre (1984), Gifi (1990), Le Roux and Rouanet (2004), Murtagh
(2005), and Nishisato (2007); the recent book, by Beh and Lombardi (2014),
presents a panoramic review of CA and related methods.

Further results on TCA can be found in Choulakian et al. (2006), Choulakian
(2008a, 2008b, 2013), Choulakian and de Tibeiro (2013), Choulakian, Allard
and Simonetti (2013), Choulakian, Simonetti and Gia (2014), and
Mallet-Gauthier and Choulakian (2015).

\subsection{Nega coding for rank data}

In the sequel, we use the same notation as in Choulakian (2014). Let ${\bf R}%
=(r_{ij})$ for $i=1,...,n$ and $j=1,...,d$ represent the Borda scores for
rank data, where $r_{ij}$ takes values $0,...,d-1.$ Similarly, $\overline{%
{\bf R}}$ represent the reverse borda scores. We note that ${\bf R}$ and $%
\overline{{\bf R}}$ contain the same information. To incorporate both in one
data set, there are two approaches in correspondence analysis literature. In
the first approach we vertically concatenate both tables, that is, we double
the size of the coded data set by defining ${\bf R}_{D}^{\prime }=({\bf R}%
^{\prime }$...$\overline{{\bf R}}^{\prime }).$ In the second approach, we
summarize $\overline{{\bf R}}$ by its column total, that is, we create a row
named ${\bf nega=}$ $n\overline{{\bf \beta }}={\bf 1}_{n}^{\prime }\overline{%
{\bf R}},$ then we vertically concatenate ${\bf nega}$ to ${\bf R}$, thus
obtaining%
\[
{\bf R}_{nega}=(_{{\bf nega}}^{{\bf R}_{nega1}}),
\]%
where ${\bf R}_{nega1}={\bf R}.$ The size of ${\bf R}_{nega}$ is $%
(n+1)\times d.$ Choulakian (2014) discussed the relationship between TCA of $%
{\bf R}_{D}$ and TCA of ${\bf R}_{nega}.$ We will consider only the
application of TCA to ${\bf R}_{nega}$, because this will show if the rank
data set is globally homogenous or heterogenous. So let
\[
{\bf P}_{nega}=(_{{\bf p}_{nega}}^{{\bf P}_{nega1}})
\]%
be the correspondence table associated with ${\bf R}_{nega}.$ Note that $%
{\bf P}_{nega1}$ is a matrix of size $n\times d$ and ${\bf p}_{nega}$ is a
row vector of size $(1,d)$. We denote the sequence of principal axes, and
the associated basic vectors, TCA vectors of principal factor scores and
dispersion measures by $$({\bf u}_{\alpha }^{nega},{\bf v}_{\alpha }^{nega},%
{\bf a}_{\alpha }^{nega},{\bf b}_{\alpha }^{nega},{\bf f}_{\alpha }^{nega},%
{\bf g}_{\alpha }^{nega},\lambda _{\alpha }^{nega})$$ for $\alpha =1,...,k$
and $k=rank({\bf P}_{nega})-1$. The following theorem is fundamental, and it
relates ${\bf \beta ,}$ the average BC score of items, to the first
principal TCA factor score ${\bf g}_{1}^{nega}$.$\bigskip $

{\bf Theorem 1 (}${\bf P}_{nega}$): {\bf (}This is Theorem 2 in Choulakian
(2014){\bf \ )}: Properties a, b, c are true iff\ ${\bf v}%
_{1}^{nega}=(_{-1}^{{\bf 1}_{n}})$, where ${\bf v}_{1}^{nega}$ is the first
principal axis of the columns of ${\bf P}_{nega}.$

a) The first principal column factor score ${\bf g}_{1}^{nega}$ of the $d$
items is an affine function of the average BC score ${\bf \beta };$ that is,

\[
corr({\bf g}_{1}^{nega},{\bf \beta })=1.
\]

b) The first nontrivial TCA dispersion measure equals twice taxicab norm of
the row vector ${\bf p}_{nega}^{(1)}$

\begin{eqnarray*}
\lambda _{1}^{nega} &=&||{\bf b}_{1}^{nega}{\bf ||}_{1}, \\
&=&2\ ||{\bf p}_{nega}^{(1)}{\bf ||}_{1}.
\end{eqnarray*}

c) Consider the residual matrix ${\bf P}_{nega}^{(2)}={\bf P}_{nega}^{(1)}%
{\bf -a}_{1}^{nega}{\bf b}_{1}^{nega^{\prime }}/\lambda _{1},$ then

\begin{eqnarray*}
{\bf P}_{nega}^{(2)} &=&(_{{\bf p}_{nega}^{(2)}}^{{\bf P}_{nega1}^{(2)}}), \\
&=&(_{{\bf o}_{d}^{\prime }}^{{\bf P}_{nega1}^{(2)}});
\end{eqnarray*}%
that is, the nega row ${\bf p}_{nega}^{(2)}={\bf 0}_{d}^{\prime }$ is the
null row vector.\bigskip

Note that in Theorem 1 we have eliminated sign-indeterminacy of the first
principal axis, by fixing ${\bf v}_{1}^{nega}(nega)=-1.$

Property a implies that the first principal factor score of the items, ${\bf %
g}_{1}^{nega},$ can be interpreted as the Borda ranking of the $d$ items.
Property b shows that the nega row of ${\bf P}_{nega}^{(1)}$ accounts for
50\% of the first nontrivial taxicab dispersion $\lambda _{1}^{nega}.$
Property c shows that the residual matrix ${\bf P}_{nega}^{(2)}$ does not
contain any information on the heavyweight nega row. Properties b and c
imply that the first nontrivial factor is completely determined by the nega
row, which plays a dominant heavyweight role, see Choulakian (2008a). Such a
context in CA is discussed by Benz{\bf \'{e}}cri (1979) using asymptotic
theory, and in dual scaling by Nishisato (1984), who names it forced
classification.

The importance of applying TCA to nega coded dataset, {\bf R}$_{nega}$, and
not to the original data set {\bf R} stems from the following two
considerations: First, if there are two columns in {\bf R} such that $%
r_{ij}=\alpha \ r_{ij_{1}}$ for $j\neq j_{1}$ and $0<\alpha ,$ then the
columns $j$ and $j_{1}$ have identical profiles, and by the invariance
property of TCA they can be merged together, which will be misleading.
Second, as discussed by Choulakian (2014), the interpretation of TCA maps of
{\bf R}$_{nega}$ is based on the law of contradiction, which will be used
recursively to find the mixture components.

\subsection{The law of contradiction}

Let $S+$ be a statement and and $S-$ its negation; then the law of
contradiction states that $S+$ and $S-$ oppose each other: they can not both
hold together, see for instance Eves (1990). We shall use the law of
contradiction as a basis for the interpretation of the maps produced by TCA
of {\bf R}$_{nega}$ in the following way. First, we recall that there are $d$
items, and we represented the Borda score of an item $j$ by the voter $i,$
for $j=1,...,d$ and $i=1,...,n$, by $r_{ij}$ and its reverse Borda score by $%
\overline{r_{ij}}.$ By the law of contradiction, $r_{ij}$ and $\overline{%
r_{ij}}$ oppose each other; which in its turn also implies that $r_{ij}$ and
$nega_{j}=\sum_{i}\overline{r_{ij}}$ oppose each other because the $nega_{j}$
contains $\overline{r_{ij}},$ or they are not associated at all if $%
nega_{j}=0$. We let
\[
{\bf f}_{1}^{nega}=(_{f_{1}(nega)}^{{\bf f}_{11}})
\]%
to represent the first TCA vector of factor scores of the $(n+1)$ rows. For
the interpretation of the results by TCA of {\bf R}$_{nega}$ we can have the
following two complementary scenarios:

Scenario 1 happens when%
\begin{equation}
{\bf f}_{1}^{nega}(i)={\bf f}_{11}(i)\geq 0\text{ \ for all }i=1,...,n\text{%
\ \ and \ \ }{\bf f}_{1}^{nega}(n+1)=\text{\ }f_{1}(nega)<0;  \tag{Scen1}
\end{equation}%
then by the law of contradiction, the first principal dimension is
interpretable and it shows the opposition between the Borda scores of the
items $r_{ij}$ to their reverse Borda scores $\overline{r_{ij}}$ summarized
by $nega_{j}$. If Scenario 1 happens, then we will say that the data set is
globally homogenous, because all voters have positive first TCA factor
scores; that is, they are directionally associated because ${\bf f}%
_{11}(i)\geq 0$ for all $i=1,...,n.$ Now by Property c of Theorem 1, the
nega row disappears and do not contribute to the higher dimensions; thus the
higher dimensions will exhibit either random noise or local heterogeneities
of the voters represented by their response patterns.

Scenario 2 is the negation of Scenario 1, it corresponds to
\begin{equation}
{\bf f}_{1}^{nega}(i)=\text{\ }{\bf f}_{11}(i)<0\text{ for some }i\text{ \
and\ }{\bf f}_{1}^{nega}(n+1)=\text{\ }f_{1}(nega)<0\text{\ };  \tag{Scen2}
\end{equation}%
then the results of TCA of {\bf Y}$_{nega}$ are not interpretable by the law
of contradiction: because some voters, say belonging to the subset V$_{1},$
are directionally associated with the nega; so to obtain interpretable
results as described in Scenario 1, we eliminate the subset of voters V$_{1}$%
, and repeat the analysis till we obtain Scenario 1. If the number of
deleted voters in V$_{1}$ is small, we consider them outliers; otherwise,
they constitute another group(s) of voters.

We have the following\bigskip

{\bf Definition 1}: If Scen1 holds, then we name the rank data {\bf R} or
{\bf R}$_{nega}$ globally homogenous.\bigskip

{\it It is of basic importance\ to note that using R}$_{nega}${\it , only
globally homogenous data are interpretable by the law of contradiction.}

\section{Global homogeneity coefficient GHC}

Rank data is much more structured than ratings data; and this aspect will be
used to propose a global homegeneity coefficient (GHC) of rank data. We
recall that
\begin{eqnarray*}
{\bf R}_{nega} &=&(_{{\bf nega}}^{{\bf R}}) \\
&=&(_{{\bf nega}}^{{\bf R}_{nega1}}),
\end{eqnarray*}%
is the nega coded rank data and
\[
{\bf P}_{nega}=(_{{\bf p}_{nega}}^{{\bf P}_{nega1}})
\]%
its associated correspondence matrix. We note the following facts:

{\it Fact 1}: The row sum of the elements of ${\bf R}_{nega}$ are: ${\bf R}%
_{nega1}{\bf 1}_{d}={\bf 1}_{n}(d(d-1)/2)$ for rows $i=1,...,n$ and $%
nd(d-1)/2$ for the nega row (or $(n+1)$th row). From which we get the total
sum of elements of ${\bf R}_{nega}$ to be ${\bf 1}_{n+1}^{\prime }{\bf R}%
_{nega}{\bf 1}_{d}=nd(d-1)$. So, the marginal relative frequency of the $i$%
th row is $1/(2n)$ for $i=1,...,n$, and, the marginal relative frequency of
the nega row is $1/2$.

{\it Fact 2}: The column sum of the elements of ${\bf R}_{nega}$ are: ${\bf 1%
}_{n+1}^{\prime }{\bf R}_{nega}=n(d-1){\bf 1}_{d}^{\prime }$ for columns $%
j=1,...,d.$ So, the marginal relative frequency of the $j$th column is $1/d$
for $j=1,...,d.$

{\it Fact 3}: Using Facts 1 and 2, we see that the first residual matrix
\[
{\bf P}_{nega}^{(1)}=(_{{\bf p}_{nega}^{(1)}}^{{\bf P}_{nega1}^{(1)}}),
\]%
has elements of :
\begin{eqnarray}
{\bf P}_{nega}^{(1)}(i,j) &=&{\bf P}_{nega1}^{(1)}(i,j)  \nonumber \\
&=&\frac{r_{ij}}{nd(d-1)}-\frac{1}{2n}.\frac{1}{d}\text{ \ \ for }i=1,...,n%
\text{ and }j=1,...,d  \nonumber \\
&=&\frac{1}{nd(d-1)}\left[ r_{ij}-\frac{(d-1)}{2}\right]
\end{eqnarray}%
and%
\begin{eqnarray*}
{\bf P}_{nega}^{(1)}(nega,j) &=&{\bf p}_{nega}^{(1)}(j) \\
&=&\frac{n(d-1)-\sum_{i=1}^{n}r_{ij}}{nd(d-1)}-\frac{1}{2}.\frac{1}{d}\text{
\ \ \ \ for }j=1,...,d.
\end{eqnarray*}%
Equation (18) states that ${\bf P}_{nega}^{(1)}$ is row centered with
respect to average ranking $\frac{(d-1)}{2}$, because $%
\sum_{j=1}^{d}r_{ij}/d=\sum_{j=0}^{d-1}j/d=\frac{(d-1)}{2}$ for $i=1,...,n,$
and $\sum_{j=1}^{d}{\bf p}_{nega}^{(1)}(j)=0$; it is also column centered.

We have the following \bigskip

{\bf Proposition 1:} For a globally homogenous rank data, $\lambda
_{1}^{nega}\geq |f_{1}(nega)|$.\bigskip

The proofs of new results are in the appendix.\bigskip

Young (1974) presented a set of four axioms that characterize uniquely BC
rule. His Axiom 4, named {\it Faithfulness,} states that when there is only
one voter, if the relation that he uses to express his preferences is so
simple that one result seems the only reasonable one, the result of the
method must be that one. The {\it Faithfulness }axiom was the inspiration of
this section. By the invariance property of TCA, that is, merging of
identical profiles does not change the results of the data analysis, a {\it %
faithfully homogenous group }is equivalent to the existence of one response
pattern $(n=1)$, and its Borda score values for a complete linear order of $%
d $ items can be represented by ${\bf r}=(d-1\ \ d-2\ \ \ .....\ \ 1\ \ 0)$,
without loss of generality by reordering of the items. Then the nega coded
correspondence table, ${\bf P}_{nega},$ will have only two rows and $d$
columns%
\begin{eqnarray*}
{\bf P}_{nega} &=&(_{{\bf r}_{nega}}^{{\bf r}})/(d(d-1)) \\
&=&(_{\overline{{\bf r}}}^{{\bf r}})/(d(d-1)),
\end{eqnarray*}%
and ${\bf P}_{nega}^{(1)}$ will be of rank 1; that is, there will be only
one principle factor, for which we note its taxicab dispersion measure by $%
U(d)$ for a fixed finite integer value of $d=2,3,....$The following result
gives the value of $U(d)$ explicitly.\bigskip

{\bf Theorem 2 (Faithfully homogenous group)}:

a) For $d=2m$ or $2m-1$ for $m\in
\mathbf{N}
^{+}$, then%
\begin{eqnarray*}
U(2m) &=&U(2m-1) \\
&=&\frac{m}{2m-1},
\end{eqnarray*}%
where we define $U(1)=1.$

b) The first and only factor score of the two rows are ${\bf f}%
_{1}^{nega}(2)=$ $f_{1}(nega)=-U(d)$ and ${\bf f}%
_{1}^{nega}(1)=f_{11}(1)=U(d).$

c) The first and only factor score of the $j$th item is ${\bf g}%
_{1}^{nega}(j)=\frac{d-2j+1}{d-1}$\ \ \ \ \ \ \ \ \ for $j=1,...,d.$ In
particular, we see that ${\bf g}_{1}^{nega}(1)=1,$ ${\bf g}%
_{1}^{nega}(d)=-1, $ and ${\bf g}_{1}^{nega}(j)$ for $j=1,...,d$ are
equispaced. So for an odd number of items $d=2m+1$ for $m\in
\mathbf{N}
^{+}$, we have ${\bf g}_{1}^{nega}(m+1)=0.\bigskip $

Let $card(A)$ denote the cardinality of a set $A$, that is, the number of
elements in $A$. The next definition formalizes the partition of a set of
items obtained in Theorem 1.\bigskip

{\bf Definition 2}: For a globally homogenous rank data set, we define a
partition of a set of items $A$ to be {\it faithful} if a) For an even
number of items $d=2m$ and $m\in
\mathbf{N}
^{+},$ the first TCA axis divides the set of items $A$ into 2 blocks such
that $A=A_{+}\sqcup A_{-},$ where ${\bf \beta }(a_{k_{j}}|a_{k_{j}}\in
A_{-})<m-1/2<{\bf \beta }(a_{k_{j}}|a_{k_{j}}\in A_{+})$ and $%
card(A_{+})=card(A_{-})=m.$

b) For an odd number of items $d=2m$ $+1$ for $m\in
\mathbf{N}
^{+},$ the first TCA axis divides the set of items $A$ into 3 blocks $%
A=A_{+}\sqcup A_{0}\sqcup A_{-},$ where ${\bf \beta }(a_{k_{j}}|a_{k_{j}}\in
A_{-})<{\bf \beta }(a_{k_{j}}|a_{k_{j}}\in A_{0})=m<{\bf \beta }%
(a_{k_{j}}|a_{k_{j}}\in A_{+}),$ and, $card(A_{+})=card(A_{-})=m$ and $%
card(A_{0})=1.\bigskip $

{\bf Remarks 1}

a) The BC for a faithfully homogenous group is ${\bf \beta =r}$, and the
Pearson correlation {\it corr(}${\bf \beta ,g}_{1}^{nega})=1$ as in Theorem
1a.

b) Theorem 2 concerns only one group. Theorem 4 generalizes Theorem 2 to
multiple faithfully homogenous subgroups; however in the multiple case only
parts a and b of Theorem 2 are satisfied and not part c. The maximum number
of multiple faithfully homogenous subgroups is $(m!)^{2}$ for $d=2m$ or $%
d=2m+1$ and $m\in
\mathbf{N}
^{+},$ which represents the number of within (intra) block permutations of
the rankings.\bigskip

The next result shows that $U(d)$ is an upper bound for the first TCA
dispersion measure $\lambda _{1}^{nega}.\bigskip $

{\bf Theorem 3}: For a globally homogenous rank data set we have%
\[
\lambda _{1}^{nega}\leq U(d).
\]%
\bigskip

{\bf Definition 3:} Based on Theorem 3 we define for a globally homogenous
rank data the following global homogeneity coefficient%
\[
GHC\text{ in }\%=100\text{ }\lambda _{1}^{nega}/U(d).
\]

$GHC$ takes values between 0 and 100. In real applications we seldom find
the value of $GHC=100\%$. However, it may approach 100\% as in the Potato's
rank data set considered later on.\bigskip

{\bf Theorem 4}: $GHC=100\%$ if and only if there is a faithful partition of
the items and the Borda scores of all voters are intra (within) block
permutations.\bigskip

{\bf Corollary 1}: $GHC=100\%$ if and only if $U(d)=\lambda
_{1}^{nega}=-f_{1}(nega)=$ ${\bf f}_{11}(i)$ \ for $i=1,...,n$.\bigskip

The following result complements Proposition 1.\bigskip

{\bf Corollary 2:} For a globally homogenous rank data, $U(d)\geq |{\bf f}%
_{11}(i)|$ for $i=1,...,n$.\bigskip

{\bf Definition 4}: A voter is named faithful if its first factor score $%
{\bf f}_{11}(voter)=U(d).$\bigskip

In the next subsection we explain these results.

\subsection{Interpretation of GHC}

\begin{itemize}
\item We consider the following artificial example with two voters and eight
items.
\end{itemize}

\begin{tabular}{|l||r|r|r|r|r|r|r|r|}
\hline
& \multicolumn{8}{||c|}{items} \\ \cline{2-9}
& {\it A} & {\it B} & {\it C} & {\it D} & {\bf E} & {\bf F} & {\bf G} & {\bf %
H} \\ \hline
voter 1 & {\it 7} & {\it 6} & {\it 5} & {\it 4} & {\bf 3} & {\bf 2} & {\bf 1}
& {\bf 0} \\
voter 2 & {\it 4} & {\it 5} & {\it 6} & {\it 7} & {\bf 3} & {\bf 1} & {\bf 0}
& {\bf 2} \\
nega & 3 & 3 & 3 & 3 & 8 & 11 & 13 & 12 \\ \hline
${\bf \beta }$ & 5.5 & 5.5 & 5.5 & 5.5 & 3 & 1.5 & 0.5 & 1 \\ \hline
\end{tabular}

We have: $U(8)=4/7=0.5714=\lambda _{1}^{nega}=-f_{1}(nega)=$ ${\bf f}%
_{11}(1)={\bf f}_{11}(2)$ and $\lambda _{2}^{nega}=0.1071.$ So $GHC=100\%.$
Figure 1 displays the TCA biplot. The first axis subdivides the items into
two faithful blocks $\left\{ A,B,C,D\right\} $ and $\left\{ E,F,G,H\right\}
; $ additionaly, the ordering of the items on the first axis is given by the
Borda count ${\bf \beta ,}$ where we see that $\left\{ A,B,C,D\right\} \succ
E\succ F\succ H\succ G.$ We also note that the two voters are faithful and
their rankings are intra block permutations (in italics and in bold). The
second axis will represent local heterogeneity by opposing in particular
item A to item D.

\begin{figure}[ptb]
\begin{center}
\includegraphics[natheight=3.1972in, natwidth=4.2575in, height=3.2422in, width=4.3094in]{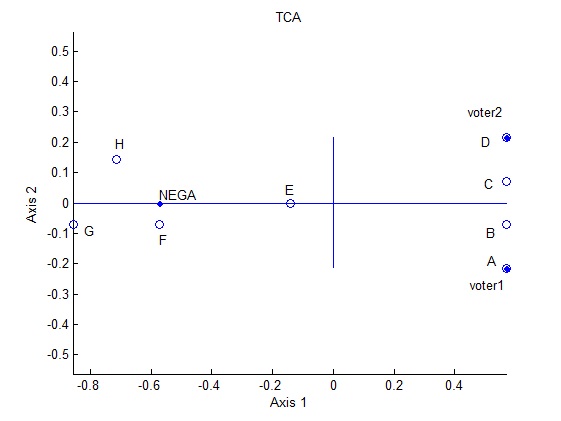}
\caption{TCA biplot of the artificial 2 voters and 8 items rank data set.}
\end{center}
\end{figure}%

\begin{itemize}
\item We consider a similar artificial example with inter block crossings
highlighted with bold characters and italics.
\end{itemize}

\begin{tabular}{|l||r|r|r|r|r|r|r|r|}
\hline
& \multicolumn{8}{||c|}{items} \\ \cline{2-9}
& {\it A} & {\it B} & {\it C} & {\it D} & {\bf E} & {\bf F} & {\bf G} & {\bf %
H} \\ \hline
voter 1 & {\it 7} & {\it 6} & {\it 5} & {\it 4} & {\bf 3} & {\bf 2} & {\bf 1}
& {\bf 0} \\
voter 2 & {\it 7} & {\it 6} & {\it 5} & {\it 4} & {\bf 3} & {\bf 2} & {\bf 1}
& {\bf 0} \\
voter 3 & {\it 7} & {\it 5} & {\it 6} & {\bf 3} & {\it 4} & {\bf 1} & {\bf 0}
& {\bf 2} \\
nega & 0 & 4 & 5 & 10 & 11 & 16 & 19 & 19 \\ \hline
${\bf \beta }$ & 21/3 & 17/3 & 16/3 & 11/3 & 10/3 & 5/3 & 2/3 & 2/3 \\ \hline
\end{tabular}

For this example we have: $U(8)=4/7=0.5714$, $\lambda _{1}^{nega}=0.5476$
and $\lambda _{2}^{nega}=0.0683.$ So $GHC=0.9583.$ Figure 2 displays the
TCA\ biplot. We note: First, the first axis subdivides the items into 2
faithful blocks $\left\{ A,B,C,D\right\} $ and $\left\{ E,F,G,H\right\} ;$
additionaly, the ordering of the items on the first axis is given by the
Borda count ${\bf \beta ,}$ where we see that $A\succ B\succ C\succ D\succ
E\succ F\succ \left\{ H,G\right\} .$ Second, voters 1 and 2 are confounded,
because they have the same profile; further $U(8)=4/7=0.5714={\bf f}_{11}(1)=%
{\bf f}_{11}(2),$ so voters 1 and 2 are faithful. Third, ${\bf f}%
_{11}(3)=0.5,$ which is smaller in value than $U(8)=0.5714,$ because voter 3
scores cross the two faithful blocks: score 4 has crossed the block $\left\{
A,B,C,D\right\} $ to the block $\left\{ E,F,G,H\right\} ,$ and score 3 has
crossed the block $\left\{ E,F,G,H\right\} $ to the block $\left\{
A,B,C,D\right\} $. The second axis will represent local heterogeneity or
random error.\bigskip


\begin{figure}[ptb]
\begin{center}
\includegraphics[natheight=3.1972in, natwidth=4.2575in, height=3.2422in, width=4.3094in]{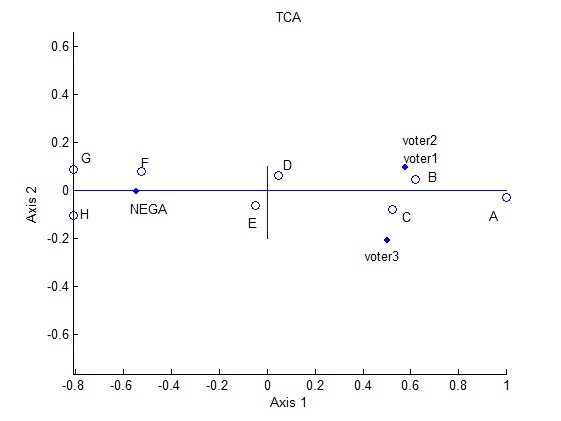}
\caption{TCAbiplot of the artificial data of 3 voters and 8 items with one crossing.}
\end{center}
\end{figure}%

\begin{itemize}
\item As an application of Theorem 4, we consider Inglehart's a priori
classification summarized in part a in Table 2. We mentioned that,
Inglehart's {\it materialist} group, composed of 4 response patterns, has $%
GHC=100\%$ and the {\it postmaterialist} group, also composed of other 4
response patterns, has also $GHC=100\%$; there is no value of GHC for the
{\it mixed} group, because the {\it mixed} group is not globally homogenous.
Let us consider Inglehart's {\it materialist} group where the pair of
materialist items $\left\{ A,C\right\} $ are always ranked above the pair of
postmaterialist items $\left\{ B,D\right\} $; from Table 1 we have (where we
permuted the positions of items C and B):%
\[
\begin{tabular}{|l||r|r|r|r|}
\hline
& \multicolumn{4}{||c|}{items} \\ \cline{2-5}
& \multicolumn{1}{||r|}{\bf A} & {\bf C} & {\it B} & {\it D} \\ \hline
A$\succ $C$\succ $B$\succ $D, 309 & \multicolumn{1}{||r|}{\bf 3} & {\bf 2} &
{\it 1} & {\it 0} \\
A$\succ $C$\succ $D$\succ $B, 255 & \multicolumn{1}{||r|}{\bf 3} & {\bf 2} &
{\it 0} & {\it 1} \\
C$\succ $A$\succ $B$\succ $D, 330 & \multicolumn{1}{||r|}{\bf 2} & {\bf 3} &
{\it 1} & {\it 0} \\
C$\succ $A$\succ $D$\succ $B, 294 & \multicolumn{1}{||r|}{\bf 2} & {\bf 3} &
{\it 1} & {\it 0} \\ \hline
${\bf \beta }$ & \multicolumn{1}{||c|}{$2.4747$} & $2.5253$ & $0.5379$ & $%
0.4621$ \\ \hline
\end{tabular}%
\]%
we see that the subsets $\left\{ A,C\right\} $ and $\left\{ B,D\right\} $
form a faithful 2 blocks partition of the four items with no crossing
between the blocks, so Theorem 4 applies and $GHC=100\%.$

\item We consider the following four orderings found in Table 1 with their
frequencies: ${\bf A}BCD137$, ${\bf A}BDC29$, ${\bf A}DBC52$ and ${\bf A}%
DCB93,$ where ABCD137 represents the ordering $A\succ B\succ C\succ D$ with
its frequency of 137. Figure 3 displays the TCA biplot.\ We can summarize
the data analysis by the following observations concerning the first axis:
First, by Definition 1 the rank data is globally homogenous, because the
factor scores of the 4 response patterns are positive on the first axis.
Second, item $A$ opposes to the items $\left\{ B,C,D\right\} ,$ so the
partition of the items is not faithful. Evidently, this means that all
voters ranked item A as their first choice; however there is considerable
heterogeneity concerning the rankings of the other 3 items $\left\{
B,C,D\right\} ;$ these local heterogeneities will appear on the second and
third axes (not shown). Third, $0.5=\lambda _{1}^{nega}=|f_{1}(nega)|={\bf f}%
_{11}(i)$ for $i=ABCD137,\ ABDC29,\ ADBC52\ $and$\ ADCB93;$ while $U(d)=2/3$
by Theorem 2$,$ so $GHC=75\%$. This example shows that the condition $%
\lambda _{1}^{nega}=|f_{1}(nega)|={\bf f}_{11}(i)$ for $i=1,...,n$ is not
sufficient for $U(d)=\lambda _{1}^{nega}.$
\end{itemize}


\begin{figure}[ptb]
\begin{center}
\includegraphics[
natheight=3.1972in, natwidth=4.2575in, height=3.2422in, width=4.3094in]{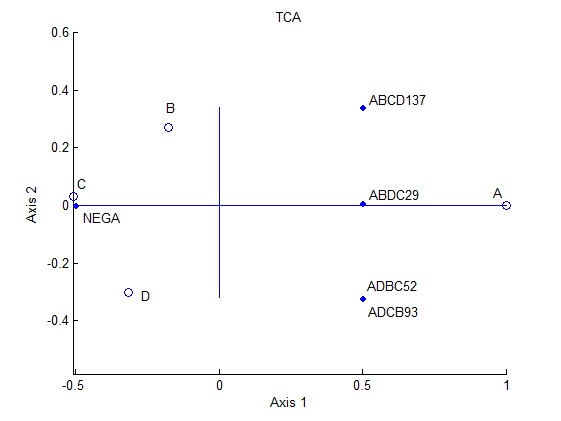}\caption{TCA
biplot of the four orderings.}
\end{center}
\end{figure}%

\bigskip

\begin{itemize}
\item The above discussion shows that: {\bf GHC takes into account the
following two aspects of rank data: a) How the items are partitioned into
blocks by the first axis. b) The extent of crossing of scores of voters
among the partitioned blocks, where the Borda count statistic provides
consensus ordering of the items on the first axis}. $GHC=100\%$ means that
the 2 or 3 blocks are faithful and for all voters their orderings of the
items are intra block permutations with no crossing between the
blocks.\bigskip
\end{itemize}

{\bf Example 1}: We consider TCA of the Potatos rank data set, found in
Vitelli et al. (2015, Table A2); it has $n=12$ assessors and $d=20$ potatos.
The first four TCA dispersion measures are: $0.5096,\ 0.0271$, $0.0219$ and $%
0.0196$; by Theorem 2, $U(20)=10/19=0.5263,$ so $GHC=100\times \frac{0.5096}{%
0.5263}=96.82\%$, which is very high. Figure 4 displays the TCA biplot of
the assessors (as points) and the true ranks of the potatos as provided in
their paper. First, on the first axis we observe a {\it faithful} partition
of the 20 potatos into 2 blocks: ten potatos T11 to T20 are found on the
left side of the first axis, and ten potatos T1 to T10 are found on the
right side of the first axis. However, potatos numbered 5 to 10 are not
correctly ranked (the true ranks of the 3 pairs $\left\{ T5,T6\right\} ,$ $%
\left\{ T7,T8\right\} $, and $\left\{ T9,T10\right\} $ are permuted). The
distribution of the majority of the true ranks of the potatos on the first
axis seem uniform. Second: the first factor score of the assessors{\bf \ }$%
{\bf f}_{11}(i)$ for $i=1,...,12$ has values $0.4947$ (4 times), $0.5053$ (2
times), $0.5158$ (3 times) and $0.5263$ (3 times), which are highly
clustered around $U(20)=10/19=0.5263$. These values show that only three
assessors are faithful; the other 9 assessors' scores have some inter block
crossings; by examining the signs of ${\bf P}_{nega1}^{(1)}$, the crossings
happened between the subsets of items $\left\{ T9,T10\right\} $ and $\left\{
T11,T12\right\} $, which are near the origin; and this is the reason that $%
GHC$ did not attain its upper value of 100\% but it approached it. Third,
given that the rank data is globally homogenous, the BC vector ${\bf \beta }$
reflects the consensus linear ordering of the potatos on the first axis,
because by Theorem 1a, $corr({\bf g}_{1}^{nega},{\bf \beta })=1.$ Fourth,
the data is almost unidimensional, because $\lambda _{2}^{nega}=0.0271$
approaches 0 and this is apparent in Figure 1. In conclusion, we can say
that this is a nice ideal real data set almost faithfully homogenous with
some random sampling error.


\begin{figure}[ptb]\begin{center}
\includegraphics[
natheight=3.1972in, natwidth=4.2575in, height=3.2422in, width=4.3094in]{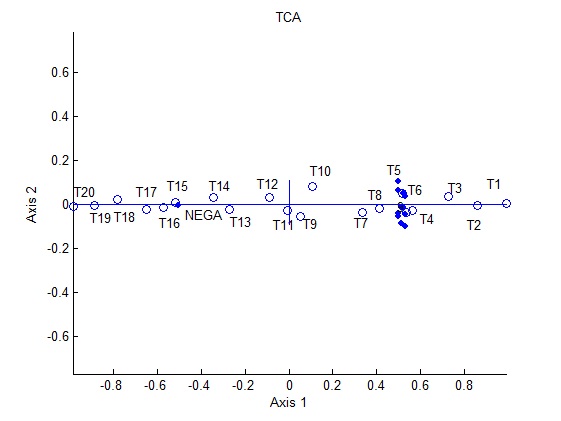}\caption{TCA
biplot of the Potatos rank data set.}
\end{center}\end{figure}%


\section{\bf Examples}

The notion of global homogeneity and local heterogeneity will further be
explained by analyzing few real data sets. First we provide some details
concerning part d of Table 2. For the other data sets we provide just
essential aspects.

\subsection{\it Example 1: Croon's political goals data continued}

Here we describe the six consecutive steps for the analysis of the rank data
set in Table 1. We recall the the description of the four items: ($A$)
maintaining order in the nation; ($B$) giving people more to say in
important government decisions; ($C$) fighting rising prices; ($D$)
protecting freedom of speech.
\begin{verbatim}
Step 1: TCA of the full data set
\end{verbatim}

Figure 5 displays the biplot of the complete data set, where to each
response pattern its observed frequency is attached; for instance the first
response pattern in Table 1, $A\succ B\succ C\succ D$ with observed
frequency of 137, is labeled as ABCD137 in the biplot. By the law of
contradiction Figure 4 is not interpretable, because there are 16 response
patterns associated with NEGA; we recall that the point NEGA contains the
negations of all response patterns. Note that the 8 response patterns $%
\left\{ BCAD61,...,DACB30\right\} $ that appear on the second axis have very
small negative values on the first axis. So we eliminate these 16 response
patterns, which have negative first factor scores, and apply TCA to the
remaining 8 response patterns in Step 2.

\begin{figure}[ptb]\begin{center}
\includegraphics[
natheight=3.1522in, natwidth=4.2575in, height=3.1964in, width=4.3094in]{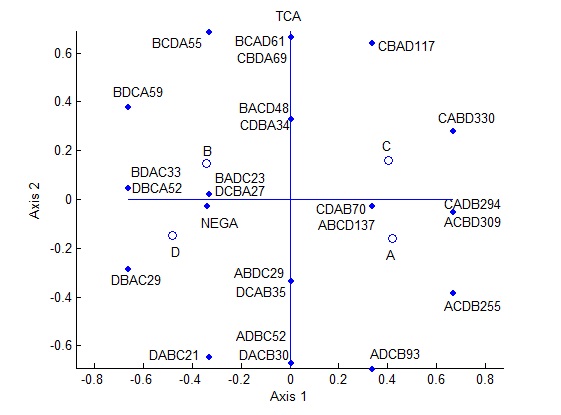}\caption{TCA
biplot of the complete Political Goals data set.}
\end{center}\end{figure}%

\begin{verbatim}
Step 2: TCA of the subset composed of 8 response patterns
\end{verbatim}

The application of TCA to the nega coded subset of weighted 8 response
patterns produces the following TCA dispersion values: $0.5801,$ $0.1127$
and $0.1116$ . Figures 6 and 7 summarize the data. Figure 6 represents the
biplot of the principal plane of the 8 response patterns; it has very clear
interpretation.

a) The first factor opposes the NEGA row to the 8 response patterns which
represent the {\it materialists }$\left\{
CDAB70,CADB294,....,ABCD137\right\} $: The 8 response patterns form a
globally homogenous group of voters, and they represent $70.95\%=(1605/2262)%
\times 100\ $of the voters in the sample; further they can be summarized by
their average BC score, ${\bf \beta }_{materialists}=(2.38,0.72,2.36,0.54),$
because by Theorem 1a, $corr({\bf \beta }_{materialists},{\bf g}%
_{1}^{nega})=1.$ Note that ${\bf g}_{1}^{nega}$ contains the first factor
coordinates of the four items plotted in Figure 6. $U(4)=2/3$ by Theorem 2,
so the global homogeneity coefficient of this group is $%
GHC(materialists)=(0.5801/U(4))\times 100=87.01\%$, which is relatively
high. On the first axis we observe the faithful partition of the items into
2 blocks, $\left\{ C,A\right\} $ and $\left\{ B,D\right\} $; but the first
factor scores of the voters have two values, $0.34$ and $0.6667=U(4)=2/3$.
This implies that the response patterns CADB294, CABD330, ACDB255 and
ACBD309 are faithful; while there are inter block crossings of scores of the
response patterns CDAB70, CBAD117, ADCB93 and ABCD137. This last assertion
is evident.

\begin{figure}[ptb]\begin{center}
\includegraphics[
natheight=3.1522in, natwidth=4.2575in, height=3.1964in, width=4.3094in]
{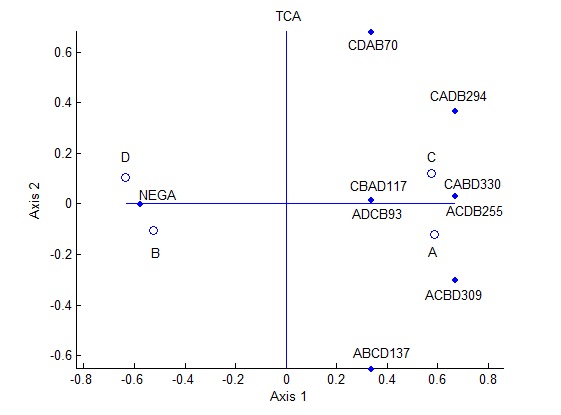}\caption{TCA
biplot of the materialists.}
\end{center}\end{figure}%


b) The NEGA point contributes only to the first axis; and by Theorem 1c, it
is eliminated from the higher axes: In Figure 7 it is found at the origin.

c) Given that the 2nd and 3rd TCA dispersion values are almost equal and
relatively high, $0.1127$ and $0.1116$, it is worthwhile to examine the
principle plane made up of axes 2 and 3, represented in Figure 7: It is
evident that there are two principle branches dominated by items A and C
respectively; these two branches represent {\it local} {\it heterogeneities}%
, in the sense that item C opposes to item A on both axes, which are both
qualified as materialist items. The two groups Postmaterialist1 and
Postmaterialist2 in Table 2, which appeared as individual groups in Croon's
SU mixture model and Lee and Yu's weighted distance based model, are the two
local branches (subdivisions) of the materialists in the TCA approach. These
two branches are similar to Marden (1995, chapter 2)'s \textquotedblright
the poles of attraction\textquotedblright\ for items A and C.

\begin{figure}[ptb]\begin{center}
\includegraphics[
natheight=3.1972in, natwidth=4.2575in, height=3.2422in, width=4.3094in]{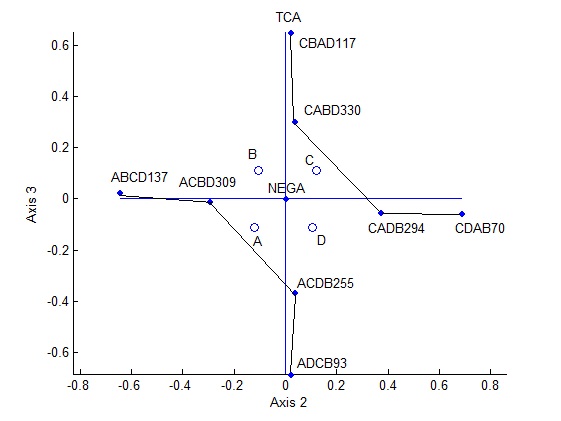}\caption{TCA
biplot showing the two branches of the materialists.}
\end{center}\end{figure}%

\begin{verbatim}
Step 3: TCA of the 16 response patterns
\end{verbatim}

We apply TCA to the 16 remaining response patterns that were associated with
the NEGA point in Figure 5, and we get Figure 8, which by the law of
contradiction is not interpretable: So we eliminate the 6 response patterns,
which are associated with the NEGA point in Figure 8; and apply TCA to the
remaining 10 response patterns in Step 4 (Step 4 is similar to Step 2).

\begin{figure}[ptb]\begin{center}
\includegraphics[
natheight=3.1972in, natwidth=4.2575in, height=3.2422in, width=4.3094in]
{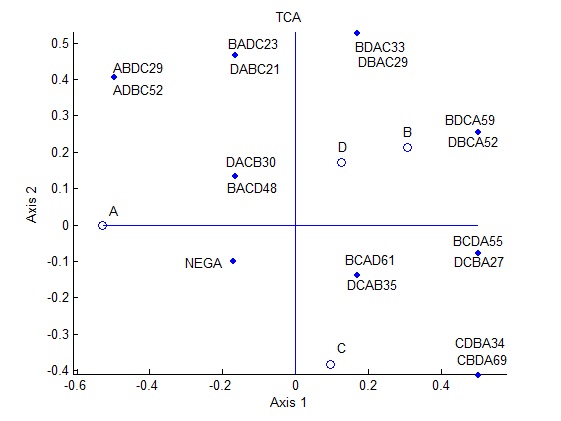}\caption{TCA
biplot of the 16 remaining response patterns.}
\end{center}\end{figure}%

\begin{verbatim}
Step 4: TCA of the 10 response patterns
\end{verbatim}

TCA dispersion measures for this case are: $0.3840,$ $0.1504$ and $0.1189$.
Figures 9 and 10 totally reflect the data. Figure 9 has the following
interpretation: a) The first factor represents the {\it postmaterialists }%
with{\it \ }$GHC=58.01\%,$ which is low for the following two reasons: the
partition of the four items into two blocks, $A$ and $\left\{ B,C,D\right\}
, $ is unfaithful by the first axis and there are a lot of inter-block
crossings by the response patterns $$\left\{
DBAC29,DCAB35,BDAC33,BCAD61\right\} $$.

\begin{figure}[ptb]\begin{center}
\includegraphics[
natheight=3.1972in, natwidth=4.2575in, height=3.2422in, width=4.3094in]
{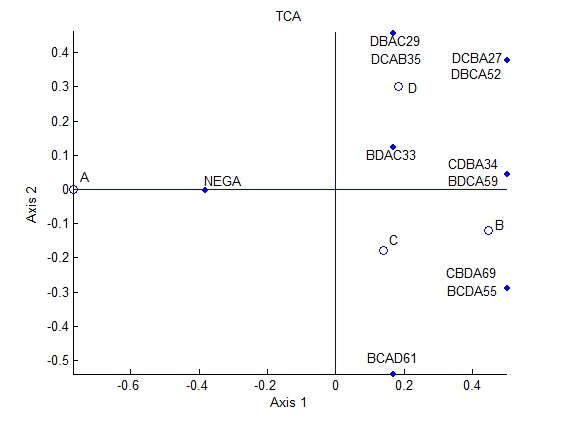}\caption{TCA
biplot of the postmaterialists. }
\end{center}\end{figure}%


b) In Figure 10, the points NEGA and item A are found on the origin: They do
not contribute to principal axes 2 and 3.

c) In Figure 10 there are three principle branches dominated by items B, D
and C, respectively; these three branches represent {\it local} {\it %
heterogeneities}, but the two branches starting with B (giving people more
to say in important government decisions) and D (protecting freedom of
speech) are more important than the smaller branch starting with item C
(fighting rising prices).

\begin{figure}[ptb]\begin{center}
\includegraphics[
natheight=3.1972in, natwidth=4.2575in, height=3.2422in, width=4.3094in]
{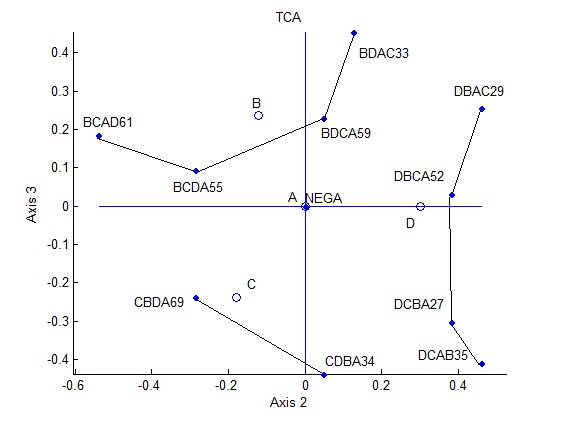}\caption{TCA
biplot showing the three branches of the postmaterialists.}
\end{center}\end{figure}%

\begin{verbatim}
Step 5: TCA of the 6 response patterns
\end{verbatim}

Figure 11 represents the TCA map of the last six patterns deleted in Step 4.
In this plot, we identify the response pattern DACB30 as an outlier because
its proportion is very small $30/2262=1.33\%$; so we eliminate it.

\begin{figure}[ptb]\begin{center}
\includegraphics[
natheight=3.1972in, natwidth=4.2575in, height=3.2422in, width=4.3094in]
{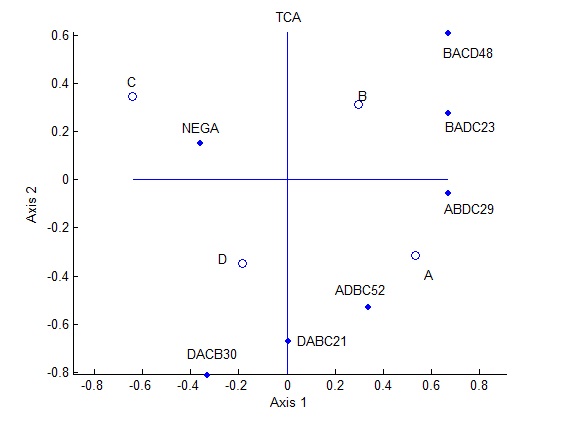}\caption{TCA
biplot of the remaining 6 response patterns.}
\end{center}\end{figure}%

\begin{verbatim}
Step 6: TCA of the remaining 5 response patterns
\end{verbatim}

TCA dispersion measures are: $0.4855,\ 0.2308$ and $0.0657$. In Figure 12
the first factor represents the {\it mixed }group. Note that on the first
axis, the mixed items $\left\{ A,B\right\} $ oppose to the mixed items $%
\left\{ D,C\right\} $; similarly, on the second axis, the mixed items $%
\left\{ B,C\right\} $ oppose to the mixed items $\left\{ A,D\right\} $.

\begin{figure}[ptb]\begin{center}
\includegraphics[
natheight=3.1972in, natwidth=4.2575in, height=3.2422in, width=4.3094in]
{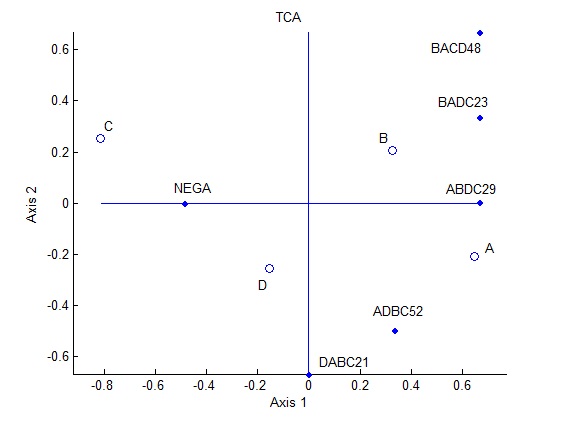}\caption{TCA
biplot of the mixeds.}
\end{center}\end{figure}%


\subsection{\it Example 1: Croon's political goals data continued with
partial ranking}

Here, we continue the analysis of Croon's political goals data by reducing
it to the first two choices, that is considering only partial rankings, as
done by Inglehart. Inglehart's approach is based on the first two choices of
the four items; thus the 24 response patterns of Table 1 is reduced to 12
response patterns. For example, the first two response patterns $A\succ
B\succ C\succ D$ and $A\succ B\succ D\succ C$ with respective frequencies
137 and 29, are collapsed into one response pattern $A\succ B\succ \ast
\succ \ast $ with frequency of 166, where * represents either C or D. Now
the Borda score of $b(A\succ B\succ \ast \succ \ast )=b$ $(A\succ B\succ
C\succ D)=(3\ \ 2\ \ 0.5\ \ 0.5),$ that is the items C and D take equal
scores. The TCA of the partial ranking table produces only two groups:
materialists $(GHC=74.8\%)$ with two branches (poles of attraction) and
postmaterialists $(GHC=52.42\%)$ with two branches. Figures 13 through 16
display these results. On these figures the label, for instance CB186,
represents the partial order $C\succ B\succ \ast \succ \ast $ with its
frequency of 186. It is obvious that there is loss of information by
reducing the complete rankings into partial rankings.

\begin{figure}[ptb]\begin{center}
\includegraphics[
natheight=3.1972in, natwidth=4.2575in, height=3.2422in, width=4.3094in]
{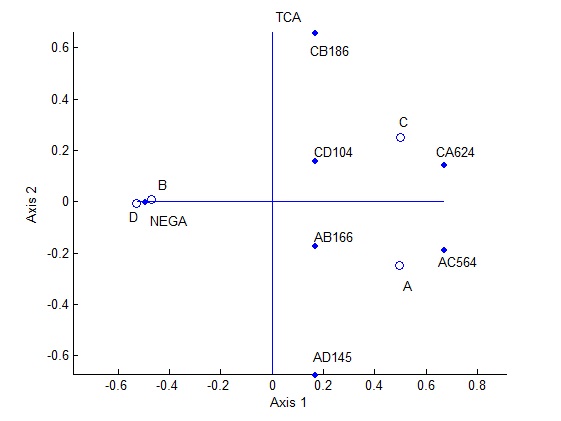}\caption{TCA
biplot of the materialists using partial rankings.}
\end{center}\end{figure}%

\begin{figure}[ptb]\begin{center}
\includegraphics[
natheight=3.1522in, natwidth=4.2575in, height=3.1964in, width=4.3094in]
{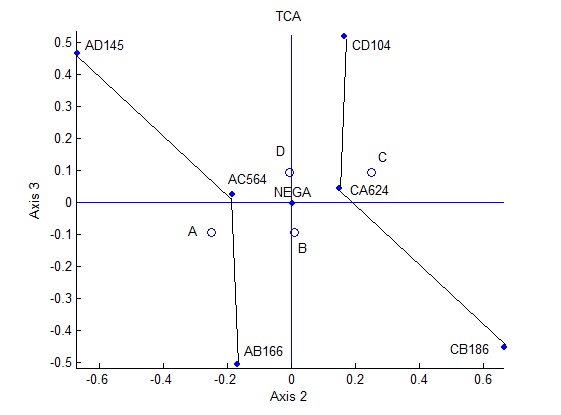}\caption{TCA
biplot showing the two branches of the materialists using partial rankings.}
\end{center}\end{figure}%

\begin{figure}[ptb]\begin{center}
\includegraphics[
natheight=3.1972in, natwidth=4.2575in, height=3.2422in, width=4.3094in]
{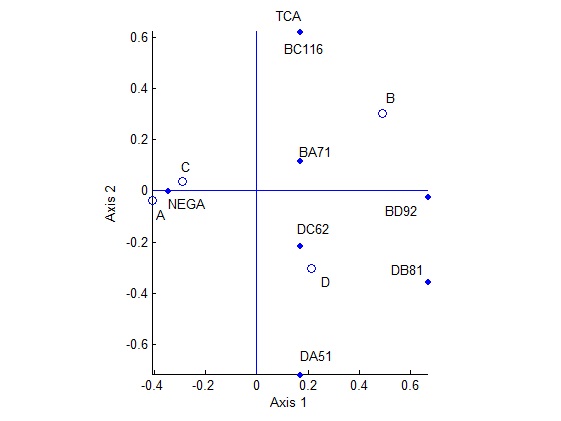}\caption{TCA
biplot of the postmaterialists using partial rankings.}
\end{center}\end{figure}%

\begin{figure}[ptb]\begin{center}
\includegraphics[
natheight=3.1972in, natwidth=4.2575in, height=3.2422in, width=4.3094in]
{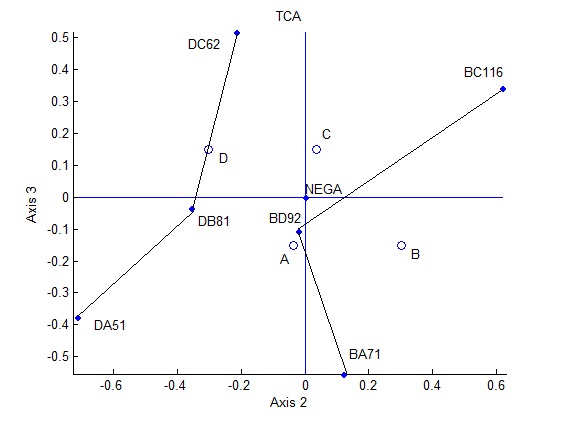}\caption{TCA
biplot showing the two branches of the postmaterialists using partial
rankings.}
\end{center}\end{figure}%

\subsection{Roskam's psychologists rank data}

Roskam preference data of size 39 by 9 was analyzed by de Leeuw (2006) and
de Leeuw and Mair (2009) and can be downloaded from their Package homals in
R. In 1968, Roskam collected preference data where 39 psychologists ranked
all 9 areas of the Psychology Department at the University of Nijmengen in
the Netherlands. The areas are: SOC = Social Psychology, EDU = Educational
and Developmental Psychology, CLI = Clinical Psychology, MAT = Mathematical
Psychology and Psychological Statistics, EXP = Experimental Psychology, CUL
= Cultural Psychology and Psychology of Religion, IND = Industrial
Psychology, TST = Test Construction and Validation, and lastly PHY =
Physiological and Animal Psychology.

de Leeuw (2006) compared linear and nonlinear principal components analysis
(PCA) approaches (Figures 4.6 and 4.7 in his paper), and concluded that
\textquotedblright the grouping in the nonlinear PCA is clearer:
psychologists in the same area are generally close together, and there is
relatively clear distinction between qualitative and quantitative
areas\textquotedblright . This assertion is true, because it describes the
two component groups of the mixture identified by TCA as will be seen.

Later on, de Leeuw and Mair (2009) applied multiple correspondence analysis,
named also homogeneity analysis, to this data set with the scale level
ordinal, and interpreted the obtained figure (Figure 8 in their paper) with
the following conclusion: \textquotedblright The plot shows interesting
rating \textquotedblright twins\textquotedblright\ of departmental areas: $%
(MAT,EXP)$,$\ (IND,TST)$, $(EDU,SOC)$,$\ (CLI,CUL)$. $\ PHY$ is somewhat
separated from the other areas\textquotedblright . This assertion does not
seem to be completely true, because of masking phenomenon.

Our TCA analysis reveals that the 39 psychologists represent a mixture of
two globally homogenous groups of sizes 23 and 16 as shown in Figures 17 and
18. In Figure 17, the following Borda ordering of the areas can be discerned
visually: $\left\{ MAT,EXP\right\} \succ \left\{ IND,TST\right\} \succ
\left\{ PHY,SOC,EDU\right\} \succ CLI\succ CUL.$ The quantitative areas of
psychology are preferred for this group. {\it Note that }$PHY${\it \ is not
separated from the rest, it has a middle ranking}.

\begin{figure}[ptb]\begin{center}
\includegraphics[
natheight=3.1522in, natwidth=4.2575in, height=3.1964in, width=4.3094in]
{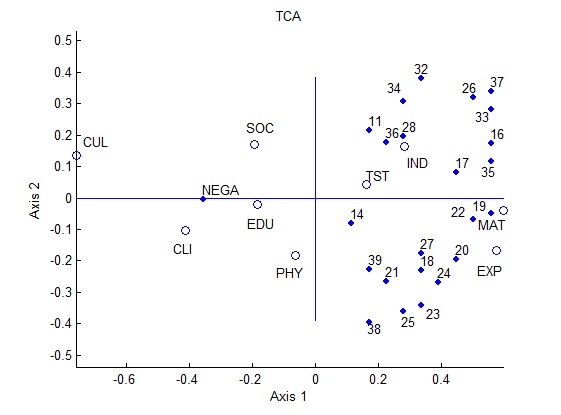}\caption{TCA
biplot of the globally homogenous group of 23 psychologists.}
\end{center}\end{figure}%

In Figure 18, the following Borda ordering of the areas can be discerned
visually: $\left\{ EDU,CLI,SOC\right\} \succ \left\{ CUL,MAT,EXP\right\}
\succ \left\{ TST,IND\right\} \succ PHY.$ For this group of psychologists
PHY is the worse rated area; further, qualitative areas are preferred for
this group.

For the complete data set, $\lambda _{1}^{nega}(complete)=0.2123$, but the
resulting TCA map is not interpretable. For group 1, $\lambda
_{1}^{nega}(group1)=0.3599$, $U(9)=5/9$, so $GHC(group1)=64.78\%$; for group
2, $\lambda _{1}^{nega}(group2)=0.3194,$ so $GHC(group2)=57.49\%.$ So,
group1 is somehat more globally homogenous than group 2; however both groups
have a lot of inter block crossings. We also note that: $\lambda
_{1}^{nega}(complete)=0.2123\ \leq \ \lambda _{1}^{nega}(group1)=0.3599$ and
$\lambda _{1}^{nega}(complete)=0.2123\ \leq \ \lambda
_{1}^{nega}(group2)=0.3194$; that is, the first TCA dispersion measure of
noninterpretable data is much smaller than the corresponding value of an
interpretable maximal subset.

\begin{figure}[ptb]\begin{center}
\includegraphics[
natheight=3.1972in, natwidth=4.2575in, height=3.2422in, width=4.3094in]
{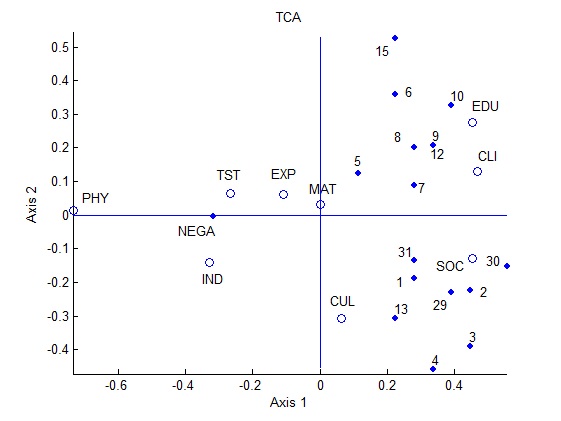}\caption{TCA
biplot of the globally homogenous group of 16 psychologists.}
\end{center}\end{figure}%

\subsection{Delbeke's family compositions rank data}

This data set of preferences can be found in Takane (2014, p.184-5): in 1978
Delbeke asked 82 Belgian university students to rank-order 16 different
family compositions, where the 16 orders are described by the coordinate
pairs $(i,j)$ for $i,j=0,...,3,$ and, the index $i$ represents the number of
daughters and the index $j$ the number of sons. This data set has been
analyzed by, among others, Heiser and de Leeuw (1981), Van Deun, Heiser and
Delbeke (2007), Takane, kiers and de leeuw (1995). In these studies, the
family composition (0,0) is considered an outlier because of its high
influence and sometimes omitted from analysis. In our approach there are no
outlier items, but voters can be tagged as outliers only by the law of
contradiction. Our results differ from theirs: We get a mixture of two
globally homogenous groups of sizes 68 and 14 as shown in Figures 19 and 20,
where points represent students and the symbol $i^{\ast }j$ represents the
family composition $(i,j)$ for $i,j=0,...,3$.

In Figure 19, we see that for this majority group of 68 students the least
preferred combination of kids is (0,0) and the most preferred combination is
(2,2). The first Borda axis opposes the combinations composed of (0
daughters or 0 sons) to the combinations composed of (at least one daughter
or at least one son). Further, we see that there is a bias towards boys: On
the first axis the position of the point $(i,j)$ is always to the left of
the point $(j,i)$. For group 1, $\lambda _{1}^{nega}(group1)=0.4017$, $%
U(16)=8/15=0.5333$, so $GHC=75.32\%$; looking at the values of the students
first factor scores, we notice 2 clusters: Cluster 1, characterized by small
number of inter blocks crossings, is composed of 26 students with first
factor score of ${\bf f}_{11}(i)=0.5250,$ 8 students with ${\bf f}%
_{11}(i)=0.5083$ and 5 students with ${\bf f}_{11}(i)=0.4917;$ this cluster,
of proportion $37/68=0.64,$ is represented by the dots making a vertical
line in Figure 15. Cluster 2, characterized by large number of inter block
crossings, are quite dispersed, having first factor scores between 0.4583
and 0.0750.

\begin{figure}[ptb]\begin{center}
\includegraphics[
natheight=3.1972in, natwidth=4.2575in, height=3.2422in, width=4.3094in]
{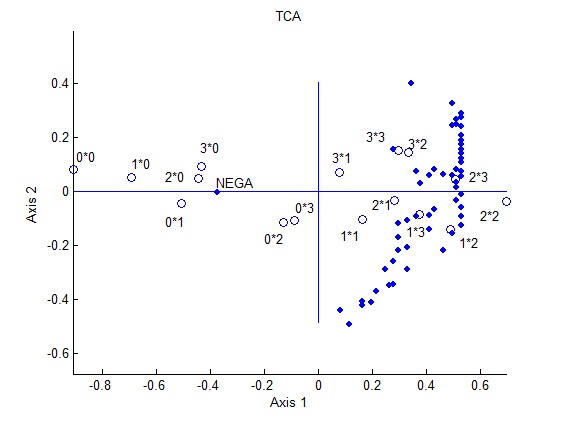}\caption{TCA
map of 68 students preferences, where i*j represents i daughters and j sons.}
\end{center}\end{figure}%


In Figure 20, for the minority group of 14 students (labeled on the biplot
by their row numbers) the least preferred combination of kids is $(3,3)$ and
the most preferred combination is $(1,1)$. The first Borda axis opposes the
combinations composed of $(0,0)$ and $(i,j)$ such that $i+j\geq 4$ to the
rest. Further, in this group also there is a bias towards boys: On the first
axis the position of the point $(i,j)$ is always to the left of the position
of the point $(j,i)$. For group 2, $\lambda _{1}^{nega}(group2)=0.4286$ and $%
GHC=80.36\%.$

\begin{figure}[ptb]\begin{center}
\includegraphics[
natheight=3.1972in, natwidth=4.2575in, height=3.2422in, width=4.3094in]
{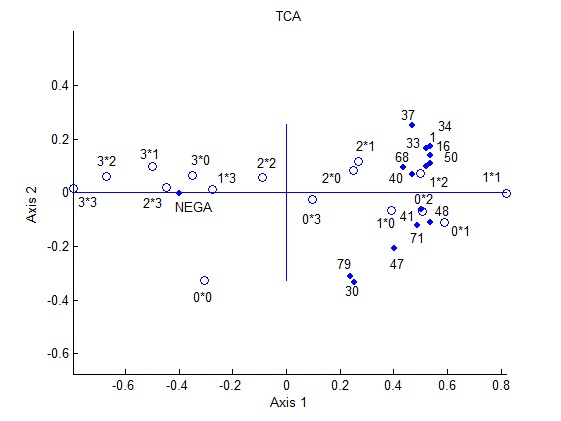}\caption{TCA
map of 14 students preferences, where i*j represents i daughters and j sons.}
\end{center}\end{figure}%

\section{\bf Conclusion}

Here we conclude with a summary of some aspects of TCA of nega coded rank
data.

We note that the rank data is spherical by nature, they are represented on a
permutahedron; so a directional method, like TCA of the nega coded rank
data, is able to discover some other aspects of the data, which are eclipsed
or masked by well established methods, such as distance or latent class
based methods. Like Occam's razor, step by step, TCA peels the essential
structural layers (globally homogenous groups) of rank data; it can also
identify outliers in a group.

We presented a new coefficient, $GHC$, that measures the global homogeneity
of a group. GHC is based on the first taxicab dispersion measure: it takes
values between 0 and 100\%, so it is easily interpretable. GHC takes into
account the following two aspects of rank data: a) How the items are
partitioned into blocks by the first axis. b) The extent of crossing of
scores of voters among the partitioned blocks, where the Borda count
statistic provides consensus ordering of the items on the first axis. $%
GHC=100\%$ means that the partition of the set of items into 2 or 3 blocks
is faithful and for all voters their orderings of the items are intra block
permutations with no crossings between the blocks. For fully ranked data,
the lower bound of $GHC$ is positive but unknown being an open problem.

As is well known, a coefficient in itself does not show important local
details in a data set. We named these important local details, local
heterogeneity; and they appear in the higher dimensions of TCA outputs: So
it is important to examine the sequence of TCA dispersion measures and the
graphical displays as expounded and professed by Benz\'{e}cri.

\bigskip

{\it Acknowleldgements: }Choulakian's research is financed by NSERC of
Canada.\bigskip

{\bf References}\medskip

Alvo, M. and Yu, P. (2014). {\it Statistical Methods for Ranking Data}.
Springer, N.Y.

Beh, E. and Lombardo, R. (2014). {\it Correspondence Analysis: Theory,
Practice and New Strategies}. N.Y: Wiley.

Benz\'{e}cri, J.P. (1966?). Sur l'analyse des pr\'{e}f\'{e}rences. ISUP
paper (available from Choulakian).

Benz\'{e}cri, J.P. (1973).\ {\it L'Analyse des Donn\'{e}es: Vol. 2:
L'Analyse des Correspondances}. Paris: Dunod.

Benz\'{e}cri, J.P. (1979). On the analysis of a table with one heavyweight
column (in french). {\it Les Cahiers de L'Analyse des Donn\'{e}es}, IV,
413-416.

Benz\'{e}cri, J.P. (1980). Geometric representation of preferences and
correspondence tables. In {\it Pratique de L'Analyse Des Donn\'{e}es}, Vol.
2, by Bastin, Ch., Benz\'{e}cri, J.P., Bourgarit, Ch. and Cazes, P. p:
299-305, Dunod, Paris.

Benz\'{e}cri, J.P (1992). {\it Correspondence Analysis Handbook}. N.Y:
Marcel Dekker.

Choulakian, V. (2006). Taxicab correspondence analysis. {\it Psychometrika,}
71, 333-345.

Choulakian, V. (2008a). Taxicab correspondence analysis of contingency
tables with one heavyweight column. {\it Psychometrika}, 73, 309-319.

Choulakian, V. (2008b). Multiple taxicab correspondence analysis. {\it %
Advances in data Analysis and Classification}, 2, 177-206.

Choulakian V. (2013). The simple sum score statistic in taxicab
correspondence analysis. In {\it Advances in Latent Variables }(ebook), eds.
Brentari E.and Carpita M., Vita e Pensiero, Milan, Italy, ISBN 978 88 343
2556 8, 6 pages.

Choulakian, V. (2014). Taxicab correspondence analysis of ratings and
rankings. {\it Journal de la Soci\'{e}t\'{e} Fran\c{c}aise de Statistique},
155(4), 1-23.

Choulakian, V. (2016). Matrix factorizations based on induced norms. {\it %
Statistics, Optimization and Information Computing}, 4, 1-14.

Choulakian, V., Allard, J. and Simonetti, B. (2013). Multiple taxicab
correspondence analysis of a survey related to health services. {\it Journal
of Data Science}, 11(2), 205-229.

Choulakian, V. and de Tibeiro, J. (2013). Graph partitioning by
correspondence analysis and taxicab correspondence analysis. {\it Journal of
Classification}, accepted to appear.

Choulakian, V., Simonetti, B. and Gia, T.P. (2014). Some new aspects of
taxicab correspondence analysis. {\it Statistical Methods and Applications},
available online.

Choulakian, V., Kasparian, S., Miyake, M., Akama, H., Makoshi, N., Nakagawa,
M. (2006). A statistical analysis of synoptic gospels. {\it JADT'2006}, pp.
281-288.

Croon, M.A. (1989). Latent class models for the analysis of rankings. In
Soete, G.D., Feger, H., Klauer, K.C. (Eds.), {\it New Developments in
Psychological Choice,} 99-121, Elsevier, Amsterdam{\it .}

de Borda, J. (1781). M\'{e}moire sur les \'{e}lections au scrutin. {\it %
Histoire de L'Acad\'{e}mie Royale des Sciences}, 102, 657-665.

de Leeuw, J (2006). Nonlinear Principal Component Analysis and Related
Techniques. In MJ Greenacre, J Blasius (eds.), {\it Multiple Correspondence
Analysis and Related Methods}, chapter 4, 107--134, Chapman \& Hall/CRC,
Boca Raton.

de Leeuw, J. and Mair, P. (2009). Homogeneity Analysis in R: the Package
homals. {\it Journal of Statistical Software}, 31(4):1-21.

Eves, H. (1997). {\it Foundations and Fundamental Concepts of Mathematics}.
N.Y. : Dover.

Fichet, B. (2009). Metrics of L$_{p}$-type and distributional equivalence
principle. {\it Advances in Data Analysis and Classification}, 3, 305-314.

Gabriel, K.R. and Zamir, S. (1979). Lower rank approximation of matrices by
least squares with any choice of weights. {\it Technometrics,} 21, 489-498.

Gifi, A. (1990). {\it Nonlinear Multivariate Analysis. }N.Y:{\it \ } Wiley.

Greenacre, M. (1984). {\it Theory and Applications of Correspondence Analysis%
}. Academic Press, London.

Heiser,W.J. and de Leeuw, J. (1981). Multidimensional mapping of preference
data. {\it Math\'{e}matiques et Sciences Humaines}, 73, 39-96.

Inglehart, R. (1977). {\it The Silent Revolution: Changing Values and
Political Styles Among Western Publics}. Princeton University Press,
Princeton.

Jacques, J., Grimonprez, Q. and Biernacki, C. (2014). RankClustr: An R
package for clustering multivariate partial rankings. {\it The R Journal},
6(1), 101-110).

Le Roux, B. and Rouanet, H. (2004). {\it Geometric Data Analysis. From
Correspondence Analysis to Structured Data Analysis}. Dordrecht:
Kluwer--Springer.

Lee, P. H. and Yu, P. L. H. (2012). Mixtures of weighted distance-based
models for ranking data with applications in political studies. {\it %
Computational Statistics and Data Analysis}, 56(8), 2486-2500.

Lee, P. and Yu, P. (2013). An R package for analyzing and modeling ranking
data. {\it BMC Medical Research Methodology}, 13(65), 1-11).

Mallet-Gauthier, S. and Choulakian, V. (2015). Taxicab correspondence
analysis of abundance data in archeology: Three case studies revisited. {\it %
Archeologia e Calcolatori}, 26, 77-94.

Marchant, T. (1998). Cardinality and the Borda score. {\it European Journal
of Operational Research}, 108, 464-472.

Marden, J.I. (1995). {\it Analyzing and Modeling of Rank Data}. Chapman \&
Hall, London, UK.

Moors, G. and Vermunt, J. (2007). Heterogeneity in post-materialists value
priorities. Evidence from a latent class discrete choice approach. {\it %
European Sociological Review,} 23(5), 631--648.

Murtagh, F. (2005). {\it Correspondence Analysis and Data Coding with Java
and R}. Boca Raton, FL., Chapman \& Hall/CRC.

Nishisato, S. (1984). Forced classification: A simple application of a
quantification method. {\it Psychometrika}, 49, 25-36.

Nishisato, S. (2007). {\it Multidimensional Nonlinear Descriptive Analysis}.
Chapman \& Hall/CRC, Baco Raton, Florida.

Pisier, G. (2012). Grothendieck's theorem, past and present. {\it Bulletin
of the American Mathematical Society}, 49 (2): 237-323.

Saari, D.G. (1990a). The Borda dictionary. {\it Social Choice and Welfare},
7, 279-317.

Saari, D.G. (1990b). Susceptibility to manipulation. {\it Public Choice,}
64, 21-41.

Saari, D.G. (1999). Explaining all three-alternative voting outcomes. {\it %
Journal of Economic Theory}, 87, 313-355.

Skrondal, A. and Rabe-Hesketh, S. (2004). {\it Generalized Latent Variable
Modeling}. Chapman \& Hall/CRC, Boca Raton, Florida.

Takane, Y. (2014). {\it Constrained Principal Compnent Analysis and Related
Techniques}. CRC Press, Bacon Raton, Florida.

Takane, Y., Kiers, H.A.L. and de Leeuw J. (1995) Component analysis with
different constraints on different dimensions. {\it Psychometrika,} 60,
259-280.

Van Deun, K., Heiser,W.J. and Delbeke, L. (2007) Multidimensional Unfolding
by Nonmetric Multidimensional Scaling of Spearman Distances in the Extended
Permutation Polytope. {\it Multivariate Behavioral Research}, 42(1),
103--132.

Vitelli, V., S\o renson, \O ., Frigessi, A. and Arjas, E. (2015).
Probabilistic preference learning with the Mallows rank model. In
arXiv:1405.7945v3.

Wold, H. (1966). Estimation of principal components and related models by
iterative least squares. In {\it Multivariate Analysis, }ed. Krishnaiah,
P.R., N.Y: Academic Press, 391-420.

Young, P. (1974). An axiomatization of Borda's rule. {\it Journal of
Economic Theory,} 9, 43-52.

\bigskip

{\bf Appendix\bigskip }

{\bf Proposition 1:} For a globally homogenous rank data, $\lambda
_{1}^{nega}\geq |f_{1}(nega)|$.

{\it Proof}: Using the same notation as in Choulakian (2014), we designate%
\[
{\bf a}_{1}^{nega}=(_{a_{1}(nega)}^{{\bf a}_{11}}).
\]%
First, by (12) we have
\begin{eqnarray*}
{\bf 1}_{n+1}^{\prime }{\bf a}_{1}^{nega} &=&0, \\
&=&{\bf 1}_{n}^{\prime }{\bf a}_{11}+a_{1}(nega);
\end{eqnarray*}%
from which we get,%
\begin{equation}
|{\bf 1}_{n}^{\prime }\ {\bf a}_{11}|=|a_{1}(nega)|.
\end{equation}%
Second, by triangle inequality of the L$_{1}$ norm we have%
\begin{equation}
||{\bf a}_{11}||_{1}\geq |{\bf 1}_{n}^{\prime }{\bf a}_{11}|.
\end{equation}

Third, by Fact 1 given in section 3 (the marginal relative frequency of the
nega row is 1/2) and (15), we have
\begin{equation}
f_{1}(nega)=2a_{1}(nega).
\end{equation}

Now we have
\begin{eqnarray*}
\lambda _{1}^{nega} &=&||{\bf a}_{1}^{nega}{\bf ||}_{1}\ \text{\ \ by (8)} \\
&=&||{\bf a}_{11}||_{1}+|a_{1}(nega)|\text{ by the additive property of L}%
_{1}\text{ norm} \\
&\geq &|{\bf 1}_{n}^{\prime }{\bf a}_{11}|+|a_{1}(nega)|\text{\ \ by (20)} \\
&=&2|a_{1}(nega)|\ \text{\ by (19)} \\
&=&|f_{1}(nega)|\text{ \ by (21).}
\end{eqnarray*}

\bigskip

{\bf Theorem 2 (Faithfully homogenous group)}:

a) For $d=2m$ or $2m-1$ and $m\in
\mathbf{N}
^{+}$, we define $U(1)=1;$ then%
\begin{eqnarray*}
U(2m) &=&U(2m-1) \\
&=&\frac{m}{2m-1}.
\end{eqnarray*}

b) The vector containing the first and only factor scores of the two rows is
${\bf f}_{1}^{nega}=(_{f_{1}(nega)}^{f_{11}(1)})=(_{-U(d))}^{U(d)}).$

c) The vector containing the first and only factor scores of the columns is $%
{\bf g}_{1}^{nega},$ where ${\bf g}_{1}^{nega}(j)=\frac{d-2j+1}{d-1}$\ \ \ \
\ \ \ \ \ for $j=1,...,d.$ In particular, we see that ${\bf g}%
_{1}^{nega}(1)=1,$ ${\bf g}_{1}^{nega}(d)=-1,$ and ${\bf g}_{1}^{nega}(j)$
for $j=1,...,d$ are equispaced. So for an odd number of items $d=2m+1$ for $%
m\in
\mathbf{N}
^{+}$, we have ${\bf g}_{1}^{nega}(m+1)=0.$

{\it Proof}: a) A {\it faithfully homogenous group }consists of one response
pattern, and its Borda score values for a complete linear order of $d$
items, without loss of generality by relabeling of the items, will be ${\bf r%
}=(d-1\ \ d-2\ \ \ .....\ \ 1\ \ 0)$. Then the nega coded correspondence
table will have only two rows and $d$ columns%
\begin{eqnarray*}
{\bf P}_{nega} &=&(_{{\bf r}_{nega}}^{{\bf r}})/(d(d-1)) \\
&=&(_{\overline{{\bf r}}}^{{\bf r}})/(d(d-1)),
\end{eqnarray*}%
and ${\bf P}_{nega}^{(1)}=(p_{ij}^{(1)}=p_{ij}-p_{i\ast }p_{\ast j})$ will
be of rank 1; that is, there is only one principle factor, for which we note
the taxicab dispersion measure by $U(d)$ for a fixed finite integer value of
$d=2,3,....$By (18) the elements of ${\bf P}_{nega}^{(1)}$ are%
\begin{eqnarray*}
{\bf P}_{nega}^{(1)}(1,j) &=&(p_{1j}-p_{1\ast }p_{\ast j}) \\
&=&(\frac{r_{1j}}{d(d-1)}-\frac{1}{2}.\frac{1}{d}) \\
&=&\frac{r_{1j}-(d-1)/2}{d(d-1)}\text{ \ \ \ \ \ for \ \ \ }j=1,...,d,
\end{eqnarray*}%
and%
\begin{eqnarray*}
{\bf P}_{nega}^{(1)}(nega,j) &=&(p_{2j}-p_{2\ast }p_{\ast j}) \\
&=&(\frac{(d-1)-r_{1j}}{d(d-1)}-\frac{1}{2}.\frac{1}{d}) \\
&=&-\frac{r_{1j}-(d-1)/2}{d(d-1)}\text{ \ \ for \ \ \ }j=1,...,d;
\end{eqnarray*}%
so%
\begin{equation}
{\bf P}_{nega}^{(1)}(1,j)=-{\bf P}_{nega}^{(1)}(nega,j).
\end{equation}%
The value of $U(d)$ by Theorem 1 b is%
\begin{eqnarray*}
U(d) &=&2\sum_{j=1}^{d}|\frac{r_{1j}-(d-1)/2}{d(d-1)}| \\
&=&\frac{1}{d(d-1)}\sum_{j=1}^{d}|d-2j+1|\text{ \ \ for }r_{1j}=d-j.
\end{eqnarray*}

\bigskip Now we consider separately even and odd values of $d$, $d=2m$ and $%
d=2m-1$ for $m\in
\mathbf{N}
^{+}$.

Case 1: $d=2m$, then
\begin{eqnarray*}
U(d &=&2m)=\frac{\sum_{j=1}^{2m}|2m-2j+1|}{2m(2m-1)} \\
&=&\frac{\sum_{j=1}^{m}(2m-2j+1)+\sum_{j=m+1}^{m}(-2m+2j-1)}{2m(2m-1)} \\
&=&\frac{2m^{2}}{2m(2m-1)} \\
&=&\frac{m}{2m-1}.
\end{eqnarray*}

Case 2: $d=2m-1$, then
\begin{eqnarray*}
U(d &=&2m-1)=\frac{\sum_{j=1}^{2m}|2m-1-2j+1|}{(2m-1)(2m-2)} \\
&=&\frac{\sum_{j=1}^{m-1}(m-j)+\sum_{j=m+1}^{2m-1}(j-m)}{(m-1)(2m-1)} \\
&=&\frac{m(m-1)}{(m-1)(2m-1)} \\
&=&\frac{m}{2m-1}.
\end{eqnarray*}

b) We have%
\begin{eqnarray*}
U(d) &=&||{\bf a}_{1}^{nega}{\bf ||}_{1}\ \text{\ \ by (8),} \\
&=&|a_{11}(1)|+|a_{1}(nega)|\text{,} \\
&=&2a_{1}(nega)\ \text{\ by (19)} \\
&=&|f_{1}(nega)|\text{ \ by (21);}
\end{eqnarray*}%
but the sign of $f_{1}(nega)$ is always negative by convention, so $%
f_{1}(nega)=-U(d)$ and $f_{11}(1)=U(d)$ by (22).

c) By (9) and (15), we have ${\bf v}_{1}^{nega}=sgn({\bf f}_{1}^{nega})=(_{%
{\bf -}1}^{1}),$ ${\bf b}_{1}^{nega}={\bf P}_{nega}^{(1)\prime }{\bf v}%
_{1}^{nega}$ and ${\bf g}_{1}^{nega}={\bf D}_{c}^{-1}{\bf b}_{1}^{nega}\ $or
elementwise
\begin{eqnarray*}
{\bf g}_{1}^{nega}(j) &=&d\text{ }{\bf b}_{1}^{nega}(j)\text{ by Fact 1
given in section 3} \\
&=&\frac{2}{(d-1)}(r_{1j}-\frac{d-1}{2}) \\
&=&\frac{2}{(d-1)}(d-j-\frac{d-1}{2}) \\
&=&\frac{d-2j+1}{d-1}\text{\ \ for}\ \ \ j=1,...,d.
\end{eqnarray*}%
\ \ \ \ \

\bigskip

{\bf Theorem 3}: For a globally homogenous rank data set we have%
\[
\lambda _{1}^{nega}\leq U(d).
\]

{\it Proof}: First, we note that ${\bf P}_{nega}^{(1)}=(_{{\bf p}%
_{nega}^{(1)}}^{{\bf P}_{nega1}^{(1)}})$ is column centered by Fact 3, so we
have%
\begin{eqnarray*}
{\bf 1}_{n+1}^{\prime }{\bf P}_{nega}^{(1)} &=&{\bf 0}_{d}^{\prime } \\
&=&{\bf 1}_{n}^{\prime }{\bf P}_{nega1}^{(1)}+{\bf p}_{nega}^{(1)};
\end{eqnarray*}%
from which we get%
\begin{equation}
||{\bf 1}_{n}^{\prime }{\bf P}_{nega1}^{(1)}||_{1}=||{\bf p}%
_{nega}^{(1)}||_{1}.
\end{equation}

Second, given that each row of ${\bf R}$ has the same values, $0,1,...,d-1$,
so we have
\begin{equation}
\sum_{j=1}^{d}|\frac{r_{1j}}{d(d-1)}-\frac{1}{2}.\frac{1}{d}|=\sum_{j=1}^{d}|%
\frac{r_{ij}}{d(d-1)}-\frac{1}{2}.\frac{1}{d}|\text{\ \ for\ }i=1,...,n.
\end{equation}

We have
\begin{eqnarray*}
\lambda _{1}^{nega} &=&2||{\bf p}_{nega}^{(1)}||_{1}\ \text{\ \ by Theorem 1b%
} \\
&=&2||{\bf 1}_{n}^{\prime }{\bf P}_{nega1}^{(1)}||_{1}\text{ by (23)} \\
&=&2\sum_{j=1}^{d}|\sum_{i=1}^{n}{\bf P}_{nega1}^{(1)}(i,j)| \\
&=&2\sum_{j=1}^{d}|\sum_{i=1}^{n}\frac{r_{ij}}{nd(d-1)}-\frac{1}{2n}.\frac{1%
}{d}|\text{ by (18)} \\
&\leq &2\sum_{j=1}^{d}\sum_{i=1}^{n}|\frac{r_{ij}}{nd(d-1)}-\frac{1}{2n}.%
\frac{1}{d}|\text{ by triangle inequality} \\
&=&2\sum_{i=1}^{n}\sum_{j=1}^{d}\frac{1}{n}|\frac{r_{1j}}{d(d-1)}-\frac{1}{2}%
.\frac{1}{d}|\text{ \ by (24)} \\
&=&2\sum_{j=1}^{d}|\frac{r_{1j}}{d(d-1)}-\frac{1}{2}.\frac{1}{d}| \\
&=&U(d).
\end{eqnarray*}

{\bf Theorem 4}: $GHI=100\%$ if and only if there is a faithful partition of
the items and the Borda scores of all voters are intra block permutations.

{\it Proof}: It is similar to the proof of Theorem 3, where the inequality
is replaced by equality. We provide a proof for $d=2m$ and $m\in
\mathbf{N}
^{+}$; for $d$ an odd integer, the proof being similar.

a) Necessary condition. We have
\begin{eqnarray*}
\lambda _{1}^{nega} &=&2||{\bf p}_{nega}^{(1)}||_{1}\ \text{\ \ by Theorem 1b%
} \\
&=&2||{\bf 1}_{n}^{\prime }{\bf P}_{nega1}^{(1)}||_{1}\text{ by (23)} \\
&=&2\sum_{j=1}^{d}|\sum_{i=1}^{n}{\bf P}_{nega1}^{(1)}(i,j)| \\
&=&\frac{2}{nd(d-1)}\sum_{j=1}^{d}|\sum_{i=1}^{n}r_{ij}-\frac{d-1}{2}|\text{
by (18)} \\
&=&\frac{2}{nd(d-1)}\sum_{i=1}^{n}\sum_{j=1}^{d}|r_{ij}-\frac{d-1}{2}| \\
&=&U(d)\text{ by (24)} \\
&=&\frac{2}{nd(d-1)}\sum_{i=1}^{n}\left[ (\sum_{j\in A_{+}}r_{ij}-\frac{d-1}{%
2})-(\sum_{j\in A_{-}}r_{ij}-\frac{d-1}{2})\right] ,
\end{eqnarray*}%
where $A_{+}=\left\{ j|\frac{d-1}{2}=m-0.5<r_{ij}\right\} $ and $%
A_{-}=\left\{ j|r_{ij}<\frac{d-1}{2}=m-0.5\right\} $ for\ $i=1,...,n.$ Given
that $\lambda _{1}^{nega}$ is a sum of $dn$ positive terms, it is easy to
see that {\bf u}$_{1}={\bf 1}_{A_{+}}-{\bf 1}_{A_{-}}$ is the first TCA
principal axis, where ${\bf 1}_{A_{+}}$ is the characteristic function of $%
{\bf 1}_{A_{+}};$ that is, it has the value of 1 if $j\in {\bf 1}_{A_{+}}$
and 0 otherwise. So for $d=2m$ and $m\in
\mathbf{N}
^{+},$ the first TCA axis divides the set of items $A$ into 2 blocks such
that $A=A_{+}\sqcup A_{-},$ with $\beta (j|j\in A_{-})<m-0.5<\beta (j|j\in
A_{+})$ and $card(A_{+})=card(A_{-})=m;$ that is, the partition of the set
of items $A$ is {\it faithful. }Furthermore,{\it \ }the Borda scores of all
voters are intra block permutations by definition of $A_{+}=\left\{ j|\frac{%
d-1}{2}=m-0.5<r_{ij}\right\} $ and $A_{-}=\left\{ j|r_{ij}<\frac{d-1}{2}%
=m-0.5\right\} $.

b) Sufficient condition. We suppose that the partition of $A=A_{+}\sqcup
A_{-}$ is faithful and there are no crossings between the 2 blocks $A_{+}$
and $A_{-};$ this implies that for $i=1,...,n:$ $r_{ij}<\frac{d-1}{2}=m-0.5$
for $j\in A_{-}$ and $\frac{d-1}{2}=m-0.5<r_{ij}$ for $j\in A_{+};$ thus we
get $U(d)=\lambda _{1}^{nega}$ as in the proof of the necessary
condition.\bigskip

{\bf Corollary 1}: $GHI=100\%$ if and only if $U(d)=\lambda
_{1}^{nega}=-f_{1}(nega)=$ ${\bf f}_{11}(i)$ \ for $i=1,...,n$.\bigskip

{\it Proof}: This follows easily from {\bf u}$_{1}={\bf 1}_{A_{+}}-{\bf 1}%
_{A_{-}}.$\bigskip

{\bf Corollary 2:} For a globally homogenous rank data, $U(d)\geq {\bf f}%
_{11}(i)$ for $i=1,...,n$.\bigskip

Proof: For an individual $i$, ${\bf f}_{11}(i)<U(d)$ happens if the
partition of the items is not faithful or there are crossings of some Borda
scores between the blocks.

\end{document}